\documentclass[preprint,amsmath,amssymb,aps]{revtex4-2}
\usepackage{graphicx}
\usepackage{dcolumn}
\usepackage{bm}
\usepackage{mathrsfs}  
\newcommand{\tens}[1]{\boldsymbol{#1}}
\newcommand{\tensunit}{\boldsymbol{1}}
\newcommand{\tprod}{\otimes}

\begin{document}
\title{Tutorial: Macroscopic QED and vacuum forces}\label{sarh:chap:macroqed}

\author{S. A. R. Horsley}
\affiliation{School of Physics and Astronomy, University of Exeter, Stocker Road, Exeter, UK, EX4 4QL}
\begin{abstract}
This tutorial introduces the theory of macroscopic QED, where a Hamiltonian is found that represents the electromagnetic field interacting with a dispersive, dissipative material.  Using a one dimensional theory as motivation, we then build up the more cumbersome three dimensional theory.  Then considering the extension of this theory to moving materials, where the material response changes due to both the Doppler effect and the mixing of electric and magnetic responses, it is shown that one gets the theory of quantum electromagnetic forces for free.  We finish by applying macroscopic QED to reproduce Pendry's expression for the quantum friction force between sliding plates.
\end{abstract}

\maketitle

\begin{quotation}
The universe was a language with a perfectly ambiguous grammar.  Every physical event was an utterance that could be parsed in two entirely different ways, one causal and the other teleological.

\noindent\emph{Ted Chiang, \textit{Story of your life and other stories, 1998}}
\end{quotation}

\section{Preliminary remarks}\label{sarh:preliminary}

\par
This tutorial describes the basics of a fully quantum mechanical approach to the theory of vacuum forces, one based upon the principle of least action.  As well as reclaiming and justifying some of the key equations of Lifshitz theory~\cite{sarh:volume9} (the workhorse of vacuum force calculations) here we shall also find some new ones, and all within a framework familiar from basic quantum mechanics: we shall work with a Hamiltonian operator and a wavefunction to describe the field, the body, and its motion.  The advantage of this approach is that it contains no assumptions about the state of the medium or the field, beyond the fact that the macroscopic Maxwell equations are valid.
\par
In the domain of Casimir physics we are in an interesting regime where we wish to calculate tiny forces on objects that are too large for us to use a microscopic theory.  Yet the force stems from an electromagnetic field with a very low amplitude, so that the description must also be quantum mechanical.  As previous chapters have indicated, in this situation we might expect the electromagnetic field to obey quantised versions of the macroscopic (spatially averaged) Maxwell equations,
\begin{equation}
	\begin{array}{ll}
		\vec{\nabla}\cdot\vec{D}&=\rho_{f}\\
		\vec{\nabla}\cdot\vec{B}&=0\\
		\vec{\nabla}\times\vec{E}&=-\frac{\partial\vec{B}}{\partial t}\\
		\vec{\nabla}\times\vec{H}&=\vec{j}_{f}+\frac{\partial\vec{D}}{\partial t}
	\end{array}
	\hspace{5mm}\longrightarrow\hspace{5mm}
	\begin{array}{ll}
		\vec{\nabla}\cdot\hat{\vec{D}}&=\hat{\rho}_{f}\\
		\vec{\nabla}\cdot\hat{\vec{B}}&=0\\
		\vec{\nabla}\times\hat{\vec{E}}&=-\frac{\partial\hat{\vec{B}}}{\partial t}\\
		\vec{\nabla}\times\hat{\vec{H}}&=\hat{j}_{f}+\frac{\partial\hat{\vec{D}}}{\partial t},
	\end{array}\label{sarh:macmax}
\end{equation}
where \(\rho_{f}\) and \(\vec{j}_{f}\) are the free charge and current density within the medium that are not induced by the field.  The set of equations on the left hand side describe the interaction of a classical electromagnetic field with a material medium (macroscopic electromagnetism), usually restricted so that the important length scales of the field are not comparable to the atomic structure of the material (See e.g. \textsection 103 in~\cite{sarh:volume8} for a discussion of macroscopic electromagnetism in cases where the scale of the field becomes comparable to the scale of the microscopic parts of the medium).  The theory implied by the right hand set of equations --- often called macroscopic quantum electrodynamics (macro--QED) --- is both macroscopic \emph{and} quantum mechanical, for the fields and sources in these equations are operators that represent spatial averages over complicated microscopic field and current distributions. Yet, to be a genuine piece of quantum physics it must be possible to derive these equations as operator equations of motion,%
\vspace{-5mm}
\begin{equation}
	\frac{\partial\hat{\vec{D}}}{\partial t} = \frac{i}{\hbar}\left[\hat{H},\hat{\vec{D}}\right]\quad
	\frac{\partial\hat{\vec{B}}}{\partial t} = \frac{i}{\hbar}\left[\hat{H},\hat{\vec{B}}\right]\label{sarh:opeqmot},
\end{equation}%
otherwise we cannot be sure that (\ref{sarh:macmax}) makes any sense.  This leads to the question: \emph{can we find the Hamiltonian that has (\ref{sarh:macmax}) as its equations of motion?}  Because electromagnetic energy is not conserved in the presence of matter the answer to this question is not obvious, even classically.  Constructing the Hamiltonian of macro--QED is the subject of the first part of this chapter.  Aspects of the theory can be found in the works of Hopfield~\cite{sarh:hopfield1958}, Huttner and Barnett~\cite{sarh:huttner1992}, Suttorp~\cite{sarh:suttorp2004}, Kheirandish and Soltani~\cite{sarh:kheirandish2008}, Scheel and Buhmann~\cite{sarh:scheel2008}, and Philbin~\cite{sarh:philbin2010}, althought this is very far from a complete list.

	The overall purpose is to show that the theory of the Casimir effect can be derived consistently from a principle of least action.  This provides a coherent understanding of both the quantum theory of light in dispersive media and quantum forces due to the electromagnetic field.  Indeed, the theory we develop from macro--QED goes beyond the results of Lifshitz theory, and can be thought of as a general quantum theory of radiation pressure.  To illustrate its utility, in the final part of the chapter we apply macro--QED to the problem of electromagnetic friction between closely spaced moving bodies (\emph{quantum friction}).
	
	The guiding principle is to describe the theory in a simple manner, which compels us to spend some time on macro--QED restricted to a single dimension.  In the end the 1D results differ very little from the three dimensional results, and the generalisation rarely involves more than performing a sum over polarisations and an integration over angles.  
\section{An introduction to macroscopic QED}\label{sarh:intromacqed}

We begin this chapter with an introduction to the simplest case of macro--QED: one polarisation of the electromagnetic field propagating in a fixed direction through a uniform dispersive medium.  We shall then illustrate the generalisation to three dimensions.

%
%
\subsection{Macroscopic QED in one dimension}\label{sarh:1dsec}
Consider an electromagnetic wave propagating along the \(x\)--axis with the electric field pointing along \(z\) and the magnetic field pointing along \(y\).  This is a special case that keeps the equations simple, but to be concrete we could imagine this to be a field in a confined geometry such as e.g. a waveguide.
%
%
%
\begin{figure}[h]
	\begin{center}
	\includegraphics[width=6.5cm]{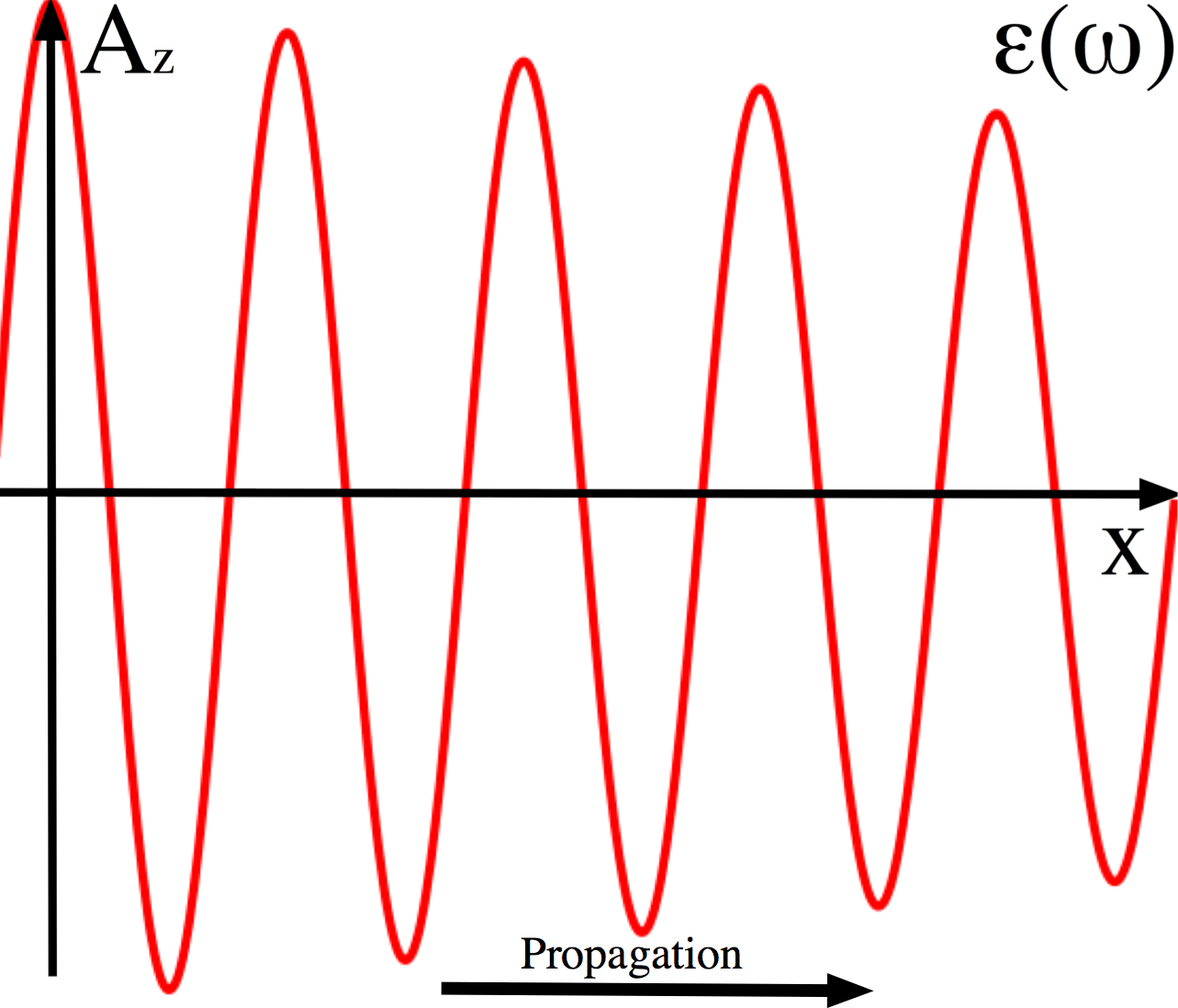}
	\caption{We consider electromagnetic waves polarised along the \(z\) axis with vector potential \(A_{z}\), and propagating along the \(x\) axis.  A material is present throughout all of space characterised by a permittivity \(\varepsilon(\omega)\), the imaginary part of which determines the rate of decay of electromagnetic waves.\label{sarh:1dwavefig}}
	\end{center}
\end{figure}
In this case there are only two non--trivial Maxwell equations,
\begin{align}
	\partial E_{z}/\partial x&=\partial B_{y}/\partial t\nonumber\\
	\partial B_{y}/\partial x&=\mu_{0}j_{z}+c^{-2}\partial E_{z}/\partial t.\label{sarh:maxwell1d}
\end{align}
Rather than working in terms of the electric and magnetic fields we write the electromagnetic field in terms of the \emph{magnetic vector potential}\index{magnetic vector potential},
\begin{equation}
	E_{z}=-\partial A_{z}/\partial t\hspace{2cm} B_{y}=-\partial A_{z}/\partial x\label{sarh:field-def}
\end{equation}
which reduces (\ref{sarh:maxwell1d}) to a single equation: the first of (\ref{sarh:maxwell1d}) is identically fulfilled and the second is the \emph{wave equation}\index{wave equation}  
\begin{equation}
	\left[\frac{\partial^{2}}{\partial x^{2}}-\frac{1}{c^{2}}\frac{\partial^{2}}{\partial t^{2}}\right]A_{z}(x,t)=-\mu_{0}j_{z}(x,t).\label{sarh:dwveq}
\end{equation}
It is thus evident that the electric current, \(j_{z}(x,t)\), is the source (or sink) for the electromagnetic waves.

\noindent\hrulefill
\paragraph*{Exercise:}  Show that if the electric field only depends on \(x\) and points along \(z\) and there is no static magnetic field, then Maxwell's equations reduce to (\ref{sarh:maxwell1d}).

\noindent\hrulefill\\

Now to introduce the medium.  We imagine that as the wave propagates through a material it slightly displaces the charges, and in doing so induces an electric current.  Such an effect can be mathematically described through writing the source on the right hand side of (\ref{sarh:dwveq}) as a linear function of the past behaviour of the electric field,
\begin{equation}
	j_{z}(x,t)=\varepsilon_{0}\int_{0}^{\infty}\chi(\tau)\frac{\partial E_{z}(x,t-\tau)}{\partial t}d\tau=-\varepsilon_{0}\int_{0}^{\infty}\chi(\tau)\frac{\partial^{2} A_{z}(x,t-\tau)}{\partial t^{2}}d\tau.\label{sarh:matsource}
\end{equation}
The function \(\chi\) is the time--dependent \emph{linear susceptibility}\index{linear susceptibility}, and represents how the effect of the electric field persists in the material over time.  The integral over \(\tau\) runs from zero to infinity because we have assumed that the medium responds to the \emph{past} behaviour of the field.

A simpler understanding of the susceptibility can be found if we write equation (\ref{sarh:dwveq}) in the frequency domain.  Writing the time dependence of the vector potential as a Fourier integral, 
\begin{equation}
	A_{z}(x,t)=\int_{-\infty}^{\infty}\frac{d\omega}{2\pi}\tilde{A}_{z}(x,\omega)e^{-i\omega t}
\end{equation}
and substituting (\ref{sarh:matsource}) into (\ref{sarh:dwveq}), gives a one dimensional Helmholtz equation for \(\tilde{A}_{z}\),
\begin{equation}
	\left[\frac{\partial^{2}}{\partial x^{2}}+\frac{\omega^{2}}{c^{2}}\varepsilon(\omega)\right]\tilde{A}_{z}(x,\omega)=0,\label{sarh:wveq}
\end{equation}
where the quantity \(\varepsilon(\omega)\) is the \emph{electric permittivity}\index{electric permittivity}.  The electric permittivity changes the wavelength within the medium from \(\lambda=2\pi c/\omega\) to \(\lambda=2\pi c/\omega\sqrt{\varepsilon}\) and is related to the time--dependent susceptibility as follows,
\begin{equation}
	\varepsilon(\omega)=1+\int_{0}^{\infty}\chi(\tau)e^{i\omega\tau}d\tau.\label{sarh:eps}
\end{equation}
From this equation we can see that it is only when the medium responds instantaneously to the field --- i.e. \(\chi(\tau)=\chi_{0}\delta(\tau)\) --- that \(\varepsilon(\omega)\) is real and frequency independent.  This is an unrealistic assumption, and in all physical cases \(\varepsilon(\omega)\) is a complex function of frequency, the imaginary part quantifying the rate at which the field is absorbed into the medium.  Due to the one--sided nature of $\chi(\tau)$---equalling zero for all times in the future--the real and imaginary parts of (\ref{sarh:eps}) are necessarily connected to one another through the \emph{Kramers--Kronig} relations,
\begin{equation}
	{\rm Re}[\varepsilon(\omega)]-1=\frac{2}{\pi}{\rm P}\int_{0}^{\infty}\frac{\omega'{\rm Im}[\varepsilon(\omega')]}{\omega'^2-\omega^2}\,d\omega'\label{eq:sarh_kk}
\end{equation}
where `$\rm P$' is Cauchy principal value of the integral.  The Kramers--Kronig relations enforce the condition that $\epsilon(\omega)$ has no poles in the upper half complex frequency plane, which is equivalent to the statement that the material responds to the past and not the future.  The interested reader can  find an excellent discussion of the Kramers--Kronig relations and the general mathematical properties of linear susceptibilities in~\cite{sarh:volume5,sarh:volume8}.

\noindent\hrulefill
\paragraph*{Exercise:}
	Show that if \(\gamma>0\) the permittivity corresponding to a single resonance at frequency \(\omega_{0}\)
	\[
		\varepsilon(\omega)=1+\frac{\omega_{P0}^{2}}{\omega_{0}^{2}-\omega^{2}-i\gamma\omega}
	\]
	does not have either poles or zeros in the upper half complex frequency plane (\(\text{Im}[\omega]>0\)).  Extend this proof to the case of \(2\) and then an arbitrary number of resonances.
	
	It can be shown (see e.g.~\cite{sarh:volume5}) that any permittivity satisfying the Kramers--Kronig relations which has a positive imaginary part is free from both poles and zeros in the upper half complex frequency plane.\label{sarh:epsexercise}
	
\noindent\hrulefill\\

The first task of this tutorial is to find an an action which yields (\ref{sarh:wveq}) as an equation of motion, for a generic permittivity that satisfies the Kramers--Kronig relations.  Then we can develop the theory of quantum light in media.  This is not as straightforward as it sounds, because the process of dissipation means that the total energy of the field is not conserved.  At the same time, the existence of a Hamiltonian (so long as it has no explicit time dependence) implies the conservation of energy.  To avoid this apparent contradiction we must introduce another system to account for the energy absorbed from the field.

%
%

\subsubsection{Mimicking a medium: finding the Lagrangian}\label{sarh:lhsec}

The aim is now as follows: find a closed system made up of the electromagnetic field plus something else, where the `something else' precisely mimics a material with complex frequency dependent permittivity, \(\varepsilon(\omega)\).  The most general way to formulate laws of motion for a closed system is to start from the \emph{principle of least action}~\cite{sarh:volume1,sarh:lanczos2013}, which is an approach that takes the actual laws of physics to be those that are optimal out of a set of possible alternatives.  Not only does this way of formulating physical theories have a deep significance, but it can be used as the basis of both classical and quantum theories of motion.

The basic quantity of interest is the action, \(S=\int L\, dt\), which is given by the integral over time of a Lagrangian, \(L\).  At a given time the Lagrangian depends on the instantaneous configuration of the system, and the classical equations of motion are obtained through finding the time evolution of the system that makes \(S\) take an extreme value.  In macroscopic electromagnetism, both the field and the material are described as continuous functions of position.  It is therefore appropriate to write the Lagrangian as the integral over space of a \emph{Lagrangian density} \(\mathscr{L}\)\footnote{For readers unfamiliar with this object see~\cite{sarh:lanczos2013,sarh:volume2}},
\begin{equation}
	L=\int\mathscr{L}dx.\label{sarh:lag}
\end{equation}

\noindent
{\footnotesize
\textbf{Example:}  Consider the case of a scalar field \(\phi(x,t)\).  The Lagrangian density is a function of the field and its derivatives, \(\mathscr{L}(\phi,\partial_{x}\phi,\partial_{t}\phi)\), and for a fixed initial and final configuration of the field, an infinitesimal change in the evolution of the field \(\phi\to\phi+\delta\phi\) induces the following change in the action
\begin{align*}
	\delta S&=\int dt\int dx\left[\delta\phi\frac{\partial\mathscr{L}}{\partial\phi}+\partial_{x}(\delta\phi)\frac{\partial\mathscr{L}}{\partial(\partial_{x}\phi)}+\partial_{t}(\delta\phi)\frac{\partial\mathscr{L}}{\partial(\partial_{t}\phi)}\right]\\
	&=\int dt\int dx\left[\frac{\partial\mathscr{L}}{\partial\phi}-\frac{\partial}{\partial x}\left(\frac{\partial\mathscr{L}}{\partial(\partial_{x}\phi)}\right)-\frac{\partial}{\partial t}\left(\frac{\partial\mathscr{L}}{\partial(\partial_{t}\phi)}\right)\right]\delta\phi,
\end{align*}
where the second line was obtained from an integration by parts.  For the action to take an extreme value with respect to all possible evolutions of the field we must at least have \(\delta S=0\), which implies
\begin{equation}
	\frac{\partial}{\partial x}\left(\frac{\partial\mathscr{L}}{\partial(\partial_{x}\phi)}\right)+\frac{\partial}{\partial t}\left(\frac{\partial\mathscr{L}}{\partial(\partial_{t}\phi)}\right)=\frac{\partial\mathscr{L}}{\partial\phi}\label{sarh:eleqns}.
\end{equation}
Equation (\ref{sarh:eleqns}) is the \emph{Euler--Lagrange}\index{Euler-Lagrange equations} field equation, which is the classical equation of motion for a field theory.  For the particular Lagrangian \(\mathscr{L}=(1/2)[c^{-2}\dot{\phi}^{2}-(\partial_{x}\phi)^{2}]\), (\ref{sarh:eleqns}) gives the wave equation without a source, \(c^{-2}\ddot{\phi}-\partial_{x}^{2}\phi=0\).\\}

\begin{figure}[h!]
	\begin{center}
	\includegraphics[width=9cm]{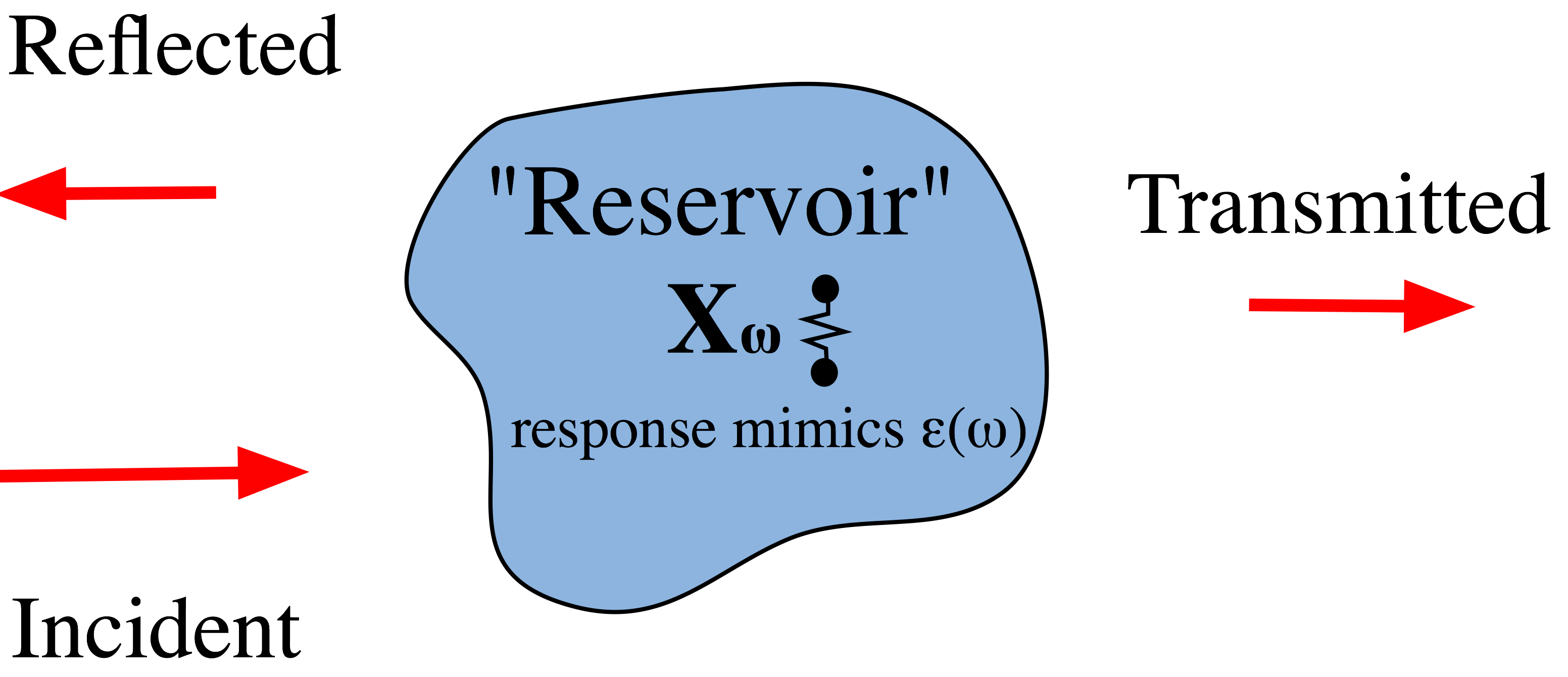}
	\caption{When electromagnetic radiation interacts with matter then some of it is inevitably absorbed: e.g. if a wave is incident onto an object then the sum of the reflected and transmitted intensities is less than that which is incident.  To model this dissipation while keeping the system closed we introduce a reservoir of simple harmonic oscillators, each with amplitude \(X_{\omega}(x,t)\).  These mimic the interaction between the field and material, and are the sink for the electromagnetic energy.\label{sarh:srfig}}
	\end{center}
\end{figure}

For the case of the electromagnetic field in a dispersive medium it is useful to consider the Lagrangian density as broken up into a sum of three parts,
\begin{equation}
	\mathscr{L}=\mathscr{L}_{F}+\mathscr{L}_{I}+\mathscr{L}_{R}\label{sarh:ld},
\end{equation}
where \(\mathscr{L}_{F}\) is the contribution due to the electromagnetic field alone, \(\mathscr{L}_{R}\) is the contribution from the system that mimics the response of the material (we shall call this the \emph{reservoir} ---Figure~\ref{sarh:srfig} shows how the reservoir ought to interact with the field), and \(\mathscr{L}_{I}\) accounts for the interaction between the two.  The part due to the field is the same as in empty space,
\begin{equation}
	\mathscr{L}_{F}=\frac{1}{2\mu_{0}}\left[\frac{1}{c^{2}}\left(\frac{\partial A_{z}(x,t)}{\partial t}\right)^{2}-\left(\frac{\partial A_{z}(x,t)}{\partial x}\right)^{2}\right],\label{sarh:field}
\end{equation}
which, as the reader can verify, reproduces the wave equation (without a source) when (\ref{sarh:eleqns}) is applied.  The reservoir (which simulates the material response, and is the sink for the electromagnetic energy) is taken as a collection of simple harmonic oscillators,
\begin{equation}
	\mathscr{L}_{R}=\frac{1}{2}\int_{0}^{\infty}\left[\left(\frac{\partial X_{\omega}(x,t)}{\partial t}\right)^{2}-\omega^{2}X_{\omega}^{2}(x,t)\right]d\omega\label{sarh:res}.
\end{equation}
After a little consideration we can see that this is a system with a tremendous number of degrees of freedom.  At each point in space there is a continuum of oscillators of amplitude \(X_{\omega}\), each labelled with a real number \(\omega\).  Specifying the instantaneous configuration of the \(X_{\omega}\) requires us to specify a function of \(\omega\) holding over the range \(\omega\in[0,\infty)\), for every point in space.  Meanwhile, \(A_{z}\) assigns just a single number to each point.  Despite the apparent application of sledgehammer to nut, the inclusion of every possible natural frequency of oscillation in (\ref{sarh:res}) is essential to reproduce the wave equation in an \emph{absorbing} material.

Each reservoir oscillator is assumed to contribute an amount \(\alpha(\omega)X_{\omega}\) to the total polarisation (dipole moment per unit volume) of the medium, where \(\alpha(\omega)\) is the material's polarisability.  Recalling the interaction energy \(-\vec{d}\cdot\vec{E}\)\index{dipole interaction} between a dipole \(\vec{d}\) and an electric field leads us to the following interaction Lagrangian
\begin{equation}
	\mathscr{L}_{I}=-\frac{\partial A_{z}(x,t)}{\partial t}\int_{0}^{\infty}\alpha(\omega)X_{\omega}(x,t)d\omega\label{sarh:coupling1}.
\end{equation}
%
%
The Lagrangian density given by the sum of (\ref{sarh:field}--\ref{sarh:coupling1}) is sufficient to reproduce the field equations in a medium (\ref{sarh:wveq}--\ref{sarh:eps}), so long as we choose the function \(\alpha(\omega)\) carefully.  Applying the Euler--Lagrange equations (\ref{sarh:eleqns}) we find the two equations of motion
\begin{align}
	\left[\frac{\partial^{2}}{\partial x^{2}}-\frac{1}{c^{2}}\frac{\partial^{2}}{\partial t^{2}}\right]A_{z}&=-\mu_{0}\frac{\partial}{\partial t}\int_{0}^{\infty}\alpha(\omega)X_{\omega}d\omega\label{sarh:fieqm}\\
	\left[\frac{\partial^{2}}{\partial t^{2}}+\omega^{2}\right]X_{\omega}&=-\alpha(\omega)\frac{\partial A_{z}}{\partial t},\label{sarh:reqm}
\end{align}
from which we can---via Eq. (\ref{sarh:dwveq})---identify the relationship between the current in the medium and the configuration of the reservoir
\begin{equation}
	j_{z}=\frac{\partial}{\partial t}\int_{0}^{\infty}\alpha(\omega)X_{\omega}d\omega\label{sarh:currres}.
\end{equation}
To mimic an absorbing medium, the reservoir must be set up in such a way that the current (\ref{sarh:currres}) is always a \emph{sink} for the field energy.

To directly compare the wave equation (\ref{sarh:fieqm}) with that in a material (\ref{sarh:wveq}), we must eliminate the reservoir from the equation for the vector potential, and thereby find the effective permittivity $\varepsilon(\omega)$.  This can be done through recognising that the above equation for \(X_{\omega}\)  (\ref{sarh:reqm}) has a general solution in terms of two real functions \(\mathcal{G}_{R,A}(t-t')\),
\begin{equation}
	\mathcal{G}_{R,A}(t-t')=\frac{1}{\omega}\begin{cases}
	\Theta(t-t')\sin[\omega(t-t')]\\
	\Theta(t'-t)\sin[\omega(t'-t)],
	\end{cases}\label{sarh:oscgreen1}
\end{equation}
which are known respectively as the \emph{retarded} and \emph{advanced} \emph{Green functions}\index{Green function} of the oscillator (\(\Theta(t-t')\) is a Heaviside step function).  These functions satisfy a differential equation very similar to (\ref{sarh:reqm}), but with a delta function on the right hand side,
\begin{equation}
	\left[\frac{\partial^{2}}{\partial t^{2}}+\omega^{2}\right]\mathcal{G}_{R,A}(t-t')=\delta(t-t')\label{sarh:oscgreen},
\end{equation}
and describe the response of the oscillator to a sudden force at \(t=t'\).  In the retarded case the oscillator responds \emph{after} the force has been applied, and in the advanced case the response occurs \emph{before} the application of the force.

\noindent\hrulefill
\paragraph*{Exercise:} Show that both of (\ref{sarh:oscgreen1}) satisfy (\ref{sarh:oscgreen}).

\noindent\hrulefill\\

Taking the retarded case \(\mathcal{G}_{R}\) and integrating it against \(\alpha(\omega)\partial A_{z}/\partial t\), we find the motion of the reservoir,
\begin{align}
	X_{\omega}(x,t)&=-\alpha(\omega)\int_{-\infty}^{\infty}\mathcal{G}_{R}(t-t')\frac{\partial A_{z}(x,t')}{\partial t'}dt'+C_{\omega}(x)e^{-i\omega t}+C^{\star}_{\omega}(x)e^{i\omega t}\nonumber\\
	&=-\frac{\alpha(\omega)}{\omega}\int_{-\infty}^{t}\sin[\omega(t-t')]\frac{\partial A_{z}(x,t')}{\partial t'}dt'+C_{\omega}(x)e^{-i\omega t}+C^{\star}_{\omega}(x)e^{i\omega t}.\label{sarh:Xsol}
\end{align}
%
%
The complex constant \(C_{\omega}(x)\) multiplies a function that satisfies the simple harmonic oscillator equation of motion (\ref{sarh:reqm}) with the right hand side equal to zero (the \emph{homogeneous} solution to the equation\index{homogeneous solution}), and is the amplitude of a current within the medium that produces radiation but is not driven by it.  In an infinite absorbing medium all radiation originates from such a current, and the state of the system can be specified entirely through the choice of \(C_{\omega}(x)\).  However, for the moment we'll set these functions to zero\footnote{This is equivalent to imposing the initial condition that \(X_{\omega}(x,t)=0\) at \(t=-\infty\)}, although in quantum mechanics these quantities play the role of creation and annihilation operators.  Substituting (\ref{sarh:Xsol}) into (\ref{sarh:fieqm}) gives the wave equation for the vector potential, now written \emph{without} reference to the oscillator amplitudes \(X_{\omega}\),
\begin{equation}
	\left[\frac{\partial^{2}}{\partial x^{2}}-\frac{1}{c^{2}}\frac{\partial^{2}}{\partial t^{2}}\right]A_{z}=\frac{1}{c^{2}}\int_{0}^{\infty}\chi(\tau)\frac{\partial^{2} A_{z}(x,t-\tau)}{\partial t^{2}}d\tau,\label{sarh:Eeqn}
\end{equation}
where we identified the susceptibility from the earlier expressions (\ref{sarh:dwveq}--\ref{sarh:matsource}), finding it equal to
\begin{equation}
	\chi(\tau)=\frac{1}{\varepsilon_{0}}\int_{0}^{\infty}\frac{\alpha^{2}(\omega)}{\omega}\sin(\omega\tau)d\omega\label{sarh:rsus}.
\end{equation}
According to (\ref{sarh:eps}), this implies the permittivity is
\begin{align}
	\varepsilon(\omega)&=1+\frac{1}{\varepsilon_{0}}\int_{0}^{\infty}\frac{\alpha^{2}(\Omega)}{\Omega}\int_{0}^{\infty}\sin(\Omega\tau)e^{i\omega\tau} d\tau d\Omega\nonumber\\
	&=1+\frac{1}{\varepsilon_{0}}\int_{0}^{\infty}\frac{\alpha^{2}(\Omega)}{2\Omega}\lim_{\eta\to0}\left[\frac{1}{\omega+\Omega+i\eta}-\frac{1}{\omega-\Omega+i\eta}\right] d\Omega\nonumber\\
	&=1+\frac{i\pi}{2\varepsilon_{0}\omega}\alpha^{2}(\omega)+\text{P}\int_{0}^{\infty}\frac{\varepsilon_{0}^{-1}\alpha^{2}(\Omega)}{\Omega^{2}-\omega^{2}} d\Omega\label{sarh:epsr}.
\end{align}
After comparison with the Kramers--Kronig relations (\ref{eq:sarh_kk}), we can see that our permittivity (\ref{sarh:epsr}) represents \emph{any} causal material with permittivity $\varepsilon(\omega)$, so long as the coupling to the \(\alpha(\omega)\) is given by
\begin{equation}
	\alpha(\omega)=\sqrt{\frac{2\omega\varepsilon_{0}\text{Im}[\varepsilon(\omega)]}{\pi}}\label{sarh:coupling},
\end{equation}
This completes the specification of our Lagrangian.  Note that, true to the spirit of the Kramers--Kronig relations, which express the connection between the real and imaginary parts of any response function, the coupling \(\alpha(\omega)\) between the field and the reservoir allows us the freedom to \emph{choose} only the imaginary part of the permittivity, the real part then emerges automatically from the equations of motion.

Our first aim is thus fulfilled: we have found a closed system that has the wave equation in an absorbing material (\ref{sarh:wveq}) as an equation of motion.  Before going any further, let us pause for a moment to consider how this theory works.  The Lagrangian (\ref{sarh:ld}) consists of the electromagnetic field coupled to an infinite number of oscillators (of every possible natural frequency) which mimic the polarisation of the material in response to the field.  There are uncountably more degrees of freedom in the reservoir than the field, so that when the reservoir is initially at rest, energy flows out of the field into the medium without coming back.  Of course real materials heat up and radiate the absorbed energy, but neglecting this is a very useful simplifying assumption of using a complex permittivity with a positive imaginary part, which is mimicked by this particular reservoir.  

\noindent\hrulefill
\paragraph*{Exercise:}  Rederive (\ref{sarh:epsr}) from the equations of motion for the field and the reservoir, but this time use the advanced Green function from (\ref{sarh:oscgreen1}).  What has happened to the permittivity?  Can you explain this?

\noindent\hrulefill\\

\subsubsection{The Hamiltonian}

The final task of this section is to find a mathematical expression for the energy of this system --- the Hamiltonian, \(H\)\index{Hamiltonian}~\cite{sarh:volume1} --- so that we can start doing some quantum mechanics.  For a field theory the Hamiltonian is given by the integral over a \emph{Hamiltonian density}\index{Hamiltonian density} \(H=\int \mathscr{H}\,dx\)~\cite{sarh:ryder2003}, which represents the energy density of the system.\\[10pt]

\noindent
{\footnotesize
\textbf{Example}:  In the case of a scalar field the Hamiltonian density is defined in terms of the Lagrangian density as follows
\begin{equation}
	\mathscr{H}=\dot{\phi}\frac{\partial\mathscr{L}}{\partial\dot{\phi}}-\mathscr{L}\equiv\dot{\phi}\Pi_{\phi}-\mathscr{L}\label{sarh:hd},
\end{equation}
where a dot above a quantity denotes a time derivative, and the \emph{canonical momentum} is given by \(\Pi_{\phi}=\partial\mathscr{L}/\partial\dot{\phi}\) (c.f. the Hamiltonian of a point particle, \(H=p\dot{x}-L\)).  Taking a partial derivative of \(\mathscr{H}\) with respect to \(\dot{\phi}\) one finds zero, which means that when we use the Hamiltonian we switch to a description in terms of the field and its canonical momentum, ceasing to use the time derivative of the field as a variable.   For the particular Lagrangian density of a free scalar field, \(\mathscr{L}=(1/2)[c^{-2}\dot{\phi}^{2}-(\partial_{x}\phi)^{2}]\), the Hamiltonian density calculated from (\ref{sarh:hd}) is
\[
	\mathscr{H}=\frac{1}{2}\left[\frac{1}{c^{2}}\dot{\phi}^{2}+\left(\frac{\partial\phi}{\partial x}\right)^{2}\right]=\frac{1}{2}\left[c^{2}\Pi_{\phi}^{2}+\left(\frac{\partial\phi}{\partial x}\right)^{2}\right].
\]\\[10pt]}
\noindent
In our case the canonical momenta are
\begin{equation}
	\Pi_{A_{z}}=\frac{\partial \mathscr{L}}{\partial \dot{A}_{z}}=\varepsilon_{0}\frac{\partial A_{z}}{\partial t}-\int_{0}^{\infty}\alpha(\omega)X_{\omega}d\omega\label{sarh:Emom}
\end{equation}
and
\begin{equation}
	\Pi_{X_{\omega}}=\frac{\partial \mathscr{L}}{\partial \dot{X}_{\omega}}=\frac{\partial X_{\omega}}{\partial t}\label{sarh:Xmom}.
\end{equation}
The two canonical momenta (\ref{sarh:Emom}--\ref{sarh:Xmom}) are evidently related to the time derivatives of the field amplitudes in quite a simple way so that it is straightforward to write the Hamiltonian density in terms of these variables:
\begin{align}
	\mathscr{H}&=\Pi_{A_{z}}\dot{A}_{z}+\int_{0}^{\infty}\Pi_{X_{\omega}}\dot{X}_{\omega}d\omega-\mathscr{L}\nonumber\\
	&=\frac{1}{2}\left[\frac{1}{\varepsilon_{0}}\left(\Pi_{A_{z}}+\int_{0}^{\infty}\alpha(\omega)X_{\omega}d\omega\right)^{2}+\frac{1}{\mu_{0}}\left(\frac{\partial A_{z}}{\partial x}\right)^{2}+\int_{0}^{\infty}\left(\Pi_{X_{\omega}}^{2}+\omega^{2}X_{\omega}^{2}\right)d\omega\right]\label{sarh:1dham}.
\end{align}
This equals the field energy plus the reservoir energy.  Although we are still working with a simplified 1D theory, this Hamiltonian is of the same form as that needed to describe the full theory of macroscopic electromagnetism~\cite{sarh:philbin2010}.

\noindent\hrulefill
\paragraph*{Exercise:} Re--express the Hamiltonian \(H=\int\mathcal{H}dx\) in terms of the electric and magnetic fields, and \(X_{\omega}\) and \(\dot{X}_{\omega}\).  What interpretation can you give the Hamiltonian when it is written in this form?

\noindent\hrulefill\\

%
%
\subsubsection{The passage from classical to quantum theory\label{sarh:1dqsec}}

Despite the fact that quantum mechanics is a conceptual break from classical physics, the formal path for constructing a quantum field theory from a classical one is straightforward in this case, and follows the textbook procedure (see for example~\cite{sarh:ryder2003}).  One route is to proceed from the expression for the action and perform a path integral~\footnote{An extensive exposition of this technique can be found in~\cite{sarh:kleinert2006}.}, and if the reader is feeling particularly keen they might want to attempt this.  However, here we take the more traditional path where the Hamiltonian is turned into an operator, and commutation relations are imposed between the fields and their canonical momenta.  The quantum mechanical version of our classical Hamiltonian (\ref{sarh:1dham}) is\index{Hamiltonian operator}
\begin{equation}
	\hat{H}=\frac{1}{2}\int d x\bigg[\frac{1}{\varepsilon_{0}}\left(\hat{\Pi}_{A_{z}}+\int_{0}^{\infty}\alpha(\omega)\hat{X}_{\omega}d\omega\right)^{2}+\frac{1}{\mu_{0}}\left(\frac{\partial \hat{A}_{z}}{\partial x}\right)^{2}+\int_{0}^{\infty}\left(\hat{\Pi}_{X_{\omega}}^{2}+\omega^{2}\hat{X}_{\omega}^{2}\right)d\omega\bigg]\label{sarh:qH},
\end{equation}
where the operators are taken to satisfy the canonical commutation relations\index{canonical commutation relations}
\begin{equation}
	\left[\hat{A}_{z}(x,t),\hat{\Pi}_{A_{z}}(x',t)\right]=i\hbar\delta(x-x')\label{sarh:com1}
\end{equation}
and
\begin{equation}
	\left[\hat{X}_{\omega}(x,t),\hat{\Pi}_{X_{\omega'}}(x',t)\right]=i\hbar\delta(\omega-\omega')\delta(x-x').\label{sarh:com2}
\end{equation}
The right hand sides of (\ref{sarh:com1}--\ref{sarh:com2}) are \(i\hbar\) times the equivalent classical Poisson brackets\footnote{A Poisson bracket\index{Poisson brackets} is a classical quantity that measures the degree of independence of the gradients of two quantities in phase space.  For example, if two quantities \(A\) and \(B\) are functions of a scalar field \(\phi\) and its canonical momentum \(\Pi_{\phi}\) the Poisson bracket is defined as, \[\left\{A,B\right\}=\int dx\bigg[\frac{\partial A}{\partial\phi(x)}\frac{\partial B}{\partial\Pi_{\phi}(x)}-\frac{\partial B}{\partial\phi(x)}\frac{\partial A}{\partial\Pi_{\phi}(x)}\bigg].\]For a full explanation of the significance of the Poisson brackets see~\cite{sarh:lanczos2013}.  For the relationship between Poisson brackets and quantum mechanics see~\cite{sarh:dirac1958}.}, which is the correspondence between classical and quantum physics that was established by Dirac~\cite{sarh:dirac1958}.  All the other commutation relations equal zero.  Equations (\ref{sarh:qH}--\ref{sarh:com2}) are the bare bones of the quantum theory of macroscopic electromagnetism, restricted to the case of a single polarisation propagating in one direction.  As described in the introduction, such a quantum theory is suitable for describing the effect of a material body (perhaps the air in this room, or a piece of metal or glass) on a very low intensity (quantum) electromagnetic field. The fact that the field amplitudes are represented by operators is significant.  

In principle we could apply (\ref{sarh:qH}) to quantum mechanical problems immediately, but at the moment it is not in a very user--friendly form.  For example, it would be some feat to directly determine the eigenstates of the system from (\ref{sarh:qH}).  Yet as this Hamiltonian contains at most quadratic combinations of the field and reservoir operators, it is not anything more than an esoteric way of writing down the Hamiltonian of a system of many coupled simple harmonic oscillators.  Therefore there is a much simpler way to write (\ref{sarh:qH}), which is to recast the system in terms of its normal modes~\footnote{See e.g.~\cite{sarh:volume1}.}.  Our case is complicated by the fact that there are infinitely many of these coupled oscillators.  Nevertheless, this transformation can be found, and one way to see what it must be is through examining the equations of motion for the operators.

\subsubsection{The operator equations of motion}

We shall now examine the behaviour of the quantum mechanical operators for the field and the medium, so as to better understand how to apply the theory of macro--QED.  From now on --- where possible --- we shall work in the Heisenberg picture\index{Heisenberg picture}\footnote{See e.g.~\cite{sarh:dirac1958}.}, where the time dependence of the system is placed in the operators rather than the wave--function.  The equations of motion for the field operators are
\begin{align}
	\frac{\partial\hat{A}_{z}}{\partial t}&=\frac{i}{\hbar}\left[\hat{H},\hat{A}_{z}\right]=\frac{1}{\varepsilon_{0}}\left(\hat{\Pi}_{A_{z}}+\int_{0}^{\infty}\alpha(\omega)\hat{X}_{\omega}d\omega\right)\nonumber\\
	\frac{\partial\hat{\Pi}_{A_{z}}}{\partial t}&=\frac{i}{\hbar}\left[\hat{H},\hat{\Pi}_{A_{z}}\right]=\frac{1}{\mu_{0}}\frac{\partial^{2}\hat{A}_{z}}{\partial x^{2}}\label{sarh:foe},
\end{align}
and those for the reservoir are
\begin{align}
	\frac{\partial\hat{X}_{\omega}}{\partial t}&=\frac{i}{\hbar}\left[\hat{H},\hat{X}_{\omega}\right]=\hat{\Pi}_{X_{\omega}}\nonumber\\
	\frac{\partial\hat{\Pi}_{X_{\omega}}}{\partial t}&=\frac{i}{\hbar}\left[\hat{H},\hat{\Pi}_{X_{\omega}}\right]=-\omega^{2}\hat{X}_{\omega}-\frac{\alpha(\omega)}{\varepsilon_{0}}\left(\hat{\Pi}_{A_{z}}+\int_{0}^{\infty}\alpha(\omega')\hat{X}_{\omega'}d\omega'\right)\label{sarh:roe},
\end{align}
both of which are, after the elimination of the canonical momenta, formally identical to the classical equations of motion (\ref{sarh:fieqm}--\ref{sarh:reqm})
\begin{align}
	\left[\frac{\partial^{2}}{\partial x^{2}}-\frac{1}{c^{2}}\frac{\partial^{2}}{\partial t^{2}}\right]\hat{A}_{z}&=-\mu_{0}\frac{\partial}{\partial t}\int_{0}^{\infty}\alpha(\omega)\hat{X}_{\omega}d\omega\label{sarh:opeq1}\\
	\left[\frac{\partial^{2}}{\partial t^{2}}+\omega^{2}\right]\hat{X}_{\omega}&=-\alpha(\omega)\frac{\partial\hat{A}_{z}}{\partial t}.\label{sarh:opeq2}
\end{align}
For the purposes of overall coherence it is worth noting that in this simplified case we have now found the mathematical route between the left and right hand sides of our earlier equation (\ref{sarh:macmax}) where we imagined the macroscopic Maxwell equations `wiht hats on': these are the macroscopic Maxwell equations.  The operator expressions that satisfy (\ref{sarh:opeq1}--\ref{sarh:opeq2}) are simply the classical expressions, but with the unknown amplitudes --- i.e. the complex quantities \(C_{\omega}(x)\) in (\ref{sarh:Xsol}) --- becoming operators.

As the classical motion of the reservoir (\ref{sarh:Xsol}) obeys an equation that is formally identical to the operator equation (\ref{sarh:opeq2}), the expression for the operator is therefore exactly the same,
\begin{equation}
	\hat{X}_{\omega}(x,t)=-\frac{\alpha(\omega)}{\omega}\int_{-\infty}^{t}\sin[\omega(t-t')]\frac{\partial \hat{A}_{z}(x,t')}{\partial t'}dt'+N_{\omega}\left[\hat{C}_{\omega}(x)e^{-i\omega t}+\hat{C}_{\omega}^{\dagger}(x)e^{i\omega t}\right]\label{sarh:Xop}.
\end{equation}
The part of the classical motion (\ref{sarh:Xsol}) that was specified by the complex amplitude \(C_{\omega}(x)\) has been replaced with the non--Hermitian operator \(\hat{C}_{\omega}(x)\), which is the annihilation operator for excitations of current within the medium (its Hermitian adjoint is the creation operator)\index{creation and annihilation operators}.  The real constant \(N_{\omega}\) is for the moment undetermined, and shall be fixed by the commutation relations (\ref{sarh:com1}--\ref{sarh:com2}).  To find a representation of the vector potential operator \(\hat{A}_{z}\) in terms of the creation and annihilation operators, (\ref{sarh:Xop}) is inserted into (\ref{sarh:opeq1}) to give
\begin{multline}
	\left[\frac{\partial^{2}\hat{A}_{z}(t)}{\partial x^{2}}-\frac{1}{c^{2}}\frac{\partial^{2}\hat{A}_{z}(t)}{\partial t^{2}}-\frac{1}{c^{2}}\int_{0}^{\infty}d\tau\chi(\tau)\frac{\partial^{2}\hat{A}_{z}(t-\tau)}{\partial t^{2}}\right]\\
	=i\mu_{0}\int_{0}^{\infty}d\omega\,\alpha(\omega)N_{\omega}\omega\left[\hat{C}_{\omega}(x)e^{-i\omega t}-\hat{C}^{\dagger}_{\omega}(x)e^{i\omega t}\right]\label{sarh:eopeqm}
\end{multline}
which has the solution
\begin{equation}
	\hat{A}_{z}(x,t)=-i\mu_{0}\int_{0}^{\infty}d\omega\,\omega N_{\omega}\alpha(\omega)\int_{-\infty}^{\infty}dx'g(x-x',\omega)\hat{C}_{\omega}(x')e^{-i\omega t}+\text{h.c.}\label{sarh:eopsol},
\end{equation}
where `\(+\,\text{h.c.}\)' means that the Hermitian conjugate of the expression should be added, and
\begin{equation}
	g(x-x',\omega)=\frac{i e^{i\frac{\omega}{c}\sqrt{\varepsilon(\omega)}|x-x'|}}{2\frac{\omega}{c}\sqrt{\varepsilon(\omega)}}\label{sarh:1dgreen}.
\end{equation}
The above quantity is the retarded Green function\index{Green function} for the electromagnetic field, and satisfies the wave equation in the frequency domain with a delta function on the right hand side
\begin{equation}
	\left[\frac{\partial^{2}}{\partial x^{2}}+\frac{\omega^{2}}{c^{2}}\varepsilon(\omega)\right]g(x-x',\omega)=-\delta(x-x')\label{sarh:1dgreeneq}.
\end{equation}

\noindent\hrulefill
\paragraph{Exercise:}Show that (\ref{sarh:1dgreen}) satisfies (\ref{sarh:1dgreeneq}).

\noindent\hrulefill\\

To Eq. (\ref{sarh:eopsol}) we could have added a superposition of solutions to the wave equation in the absence of a source, \(\exp(\pm i\omega\sqrt{\varepsilon(\omega)}x/c-i\omega t)\).  When the medium is homogeneous and absorbing, these waves grow exponentially large when \(x\) goes to either plus or minus infinity.  This divergence corresponds to the fact that within an infinitely extended absorbing medium it is impossible for a monochromatic field to exist in the absence of a source (the field is being absorbed!), and therefore these waves should not be included.  Yet in general, when the medium is not homogeneous there are solutions that do not diverge at infinity (e.g. waves incident from vacuum onto an absorbing material).  We have two options; we can either consider all of space to be filled with an absorbing medium, and take free space as the limit \(\alpha(\omega)\to0\) at the end of every calculation; or we can include these extra solutions within the Green function to ensure that there is no energy lost from the system at infinity~\footnote{A discussion of this can be found in~\cite{sarh:eckhardt1982}.}.  Having made this qualification, both electromagnetic field and reservoir operators can be written entirely in terms of \(\hat{C}_{\omega}\) and \(\hat{C}^{\dagger}_{\omega}\), and in the remainder of the text we shall assume that the Green function has the appropriate behaviour at infinity.\index{Green function, boundary conditions on}

The full expression for the reservoir operators \(\hat{X}_{\omega}\) in terms of the creation and annihilation operators is
\begin{multline}
	\hat{X}_{\omega}(x,t)=\lim_{\eta\to0}\int_{0}^{\infty}d\Omega\,\int_{\infty}^{\infty}dx'N_{\Omega}\bigg[\frac{\mu_{0}\Omega^{2}\alpha(\omega)\alpha(\Omega)g(x-x',\Omega)}{(\omega+\Omega+i\eta)(\omega-\Omega-i\eta)}\\
	+\delta(\Omega-\omega)\delta(x-x')\bigg]\hat{C}_{\Omega}(x')e^{-i\Omega t}+\text{h.c.}\label{sarh:xopsol}
\end{multline}
Expressions for the remaining operators, \(\hat{\Pi}_{X_{\omega}}\) and \(\hat{\Pi}_{A_{z}}\) can be found from applying the operator equations of motion (\ref{sarh:foe}--\ref{sarh:roe}) and are
\begin{multline}
	\hat{\Pi}_{A_{z}}(x,t)=-\int_{0}^{\infty}d\omega\, N_{\omega}\alpha(\omega)\int_{-\infty}^{\infty}dx'\bigg[\frac{\omega^{2}}{c^{2}}\varepsilon(\omega)g(x-x',\omega)\\
	+\delta(x-x')\bigg]\hat{C}_{\omega}(x')e^{-i\omega t}+\text{h.c.}\label{sarh:pezopsol}
\end{multline}
and
\begin{multline}
	\hat{\Pi}_{X_{\omega}}(x,t)=-\lim_{\eta\to0}\int_{0}^{\infty}i\Omega d\Omega\int_{-\infty}^{\infty}dx'N_{\Omega}\bigg[\frac{\mu_{0}\Omega^{2}\alpha(\omega)\alpha(\Omega)g(x-x',\Omega)}{(\omega+\Omega+i\eta)(\omega-\Omega-i\eta)}\\
	+\delta(\Omega-\omega)\delta(x-x')\bigg]\hat{C}_{\Omega}(x')e^{-i\Omega t}+\text{h.c.}\label{sarh:pxopsol}
\end{multline}
The key to understanding these expressions is that the electromagnetic field within the medium originates from a current.  In Macro--QED the \(\hat{C}_{\omega}(x)\) and \(\hat{C}^{\dagger}_{\omega}(x)\) create and annihilate the quanta of this current in the material, and are the operators that replace the photon creation and annihilation operators of ordinary quantum electrodynamics (see e.g.~\cite{sarh:ryder2003} or~\cite{sarh:dirac1958}).  The quantum theory of the electromagnetic field in a dispersive and dissipative medium is one where quanta of current are treated as the fundamental objects, and there are no photons, so to speak.  Therefore when we come to discuss the Casimir effect, the description won't be anything like Casimir's original visualisation: \emph{the force will be seen to arise from the interaction of the ground state currents within the media, rather than being due to the confined electromagnetic modes between them.}

\subsubsection{Diagonalising the Hamiltonian\index{diagonalising the Hamiltonian}}

Having re--written all of the field and reservoir operators in terms of current operators \(\hat{C}_{\omega}\) and \(\hat{C}_{\omega}^{\dagger}\), we can now identify the normal modes of the system --- then we might actually make use of the Hamiltonian!  The process of reducing the Hamiltonian to its normal modes is often referred to as \emph{Fano diagonalisation} due to the similarity with a procedure used by Fano in a study of the coupling of an atomic bound state to a continuum of (ionized) excited states\cite{sarh:fano1961}.

First notice that in both field and reservoir operators, the time dependence occurs either as a factor of \(\exp(-i\omega t)\), sitting next to \(\hat{C}_{\omega}\), or as \(\exp(i\omega t)\) sitting next to \(\hat{C}_{\omega}^{\dagger}\).  As a consequence, taking a time derivative of any of the above operators is the same as making the substitution,
\[
	\hat{C}_{\omega}(x)\to-i\omega\hat{C}_{\omega}(x).
\]
But from (\ref{sarh:foe}--\ref{sarh:roe}), the commutation between the Hamiltonian and \(\hat{C}_{\omega}\) must have the same effect as a time derivative, implying,
\begin{equation}
	\frac{i}{\hbar}\left[\hat{H},\hat{C}_{\omega}\right]=-i\omega\hat{C}_{\omega}(x).\label{sarh:Ccom}
\end{equation}
If we knew the commutation relations between \(\hat{C}_{\omega}(x)\) and \(\hat{C}_{\omega}^{\dagger}(x)\), then we could use (\ref{sarh:Ccom}) to infer the expression for the Hamiltonian in terms of these operators.  What we do know is that when the coupling between the field and reservoir is turned off (i.e. the function \(\alpha(\omega)\) is set to zero) the reservoir reduces to a field of simple harmonic oscillators uncoupled from the field and each other, with commutations relations
\begin{equation}
	\left[\hat{C}_{\omega}(x),\hat{C}_{\omega'}^{\dagger}(x')\right]=\delta(\omega-\omega')\delta(x-x')\label{sarh:Ccom2}.
\end{equation}
We make the assumption that (\ref{sarh:Ccom2}) also holds when the reservoir is coupled to the field --- an assumption which is justified below --- so that the Hamiltonian consistent with (\ref{sarh:Ccom}) is given by
\begin{align}
	\hat{H}&=\frac{1}{2}\int d x\int_{0}^{\infty}d\omega\,\hbar\omega\left[\hat{C}_{\omega}(x)\hat{C}_{\omega}(x)^{\dagger}+\hat{C}_{\omega}(x)^{\dagger}\hat{C}_{\omega}(x)\right]\nonumber\\
	&=\int d x\int_{0}^{\infty}d\omega\,\hbar\omega\left[\hat{C}_{\omega}(x)^{\dagger}\hat{C}_{\omega}(x)+\frac{1}{2}\delta(x=0)\delta(\omega=0)\right]\label{sarh:diagham},
\end{align}
which is the simplified form of the Hamiltonian (\ref{sarh:qH}) we set out to find: the system of coupled fields has been reduced to a continuum of uncoupled simple harmonic oscillators.  It is a lengthy process to explicitly verify that substituting the expressions for the operators (\ref{sarh:xopsol}--\ref{sarh:pxopsol}) into the Hamiltonian gives (\ref{sarh:diagham}), but several authors have verified this and the interested reader should consult~\cite{sarh:huttner1992,sarh:suttorp2004,sarh:philbin2010,sarh:horsley2012}.

Written in these terms, the meaning of the Hamiltonian is transparent.  The integrand consists of the operator \(\hat{C}^{\dagger}_{\omega}(x)\hat{C}_{\omega}(x)\), which is analogous to the photon number operator in QED and counts how many quanta of current per unit frequency per unit volume are within the medium.  In addition to this we have the term \(\frac{1}{2}\delta(x=0)\delta(\omega=0)\) which is the (infinite!) ground state energy of the system.  This ground state contribution can be thought of as the total energy of the system that results from the irreducible `fluctuating' current within the reservoir (medium), and in this theory it is the equivalent of the infinite ground state energy that enters Casimir's calculation of the force between two perfect mirrors.
\newpage
\noindent\hrulefill
\paragraph*{Exercise:}The ground state energy of the electromagnetic field in empty space (in 3D) equals
	\[
		\int d^{3}\vec{r}\left\langle\frac{\varepsilon_{0}}{2}\hat{\vec{E}}^{2}(\vec{r})+\frac{1}{2\mu_{0}}\hat{\vec{B}}^{2}(\vec{r})\right\rangle=\int d^{3}\vec{r}\int_{0}^{\infty}\frac{\hbar\omega^{3}}{2\pi^{2}c^{3}}d\omega
	\]
	which is infinite.  However the \emph{integrand} is finite.  Meanwhile, the ground state energy of our Hamiltonian is the integral over delta functions given in (\ref{sarh:diagham}), which is infinite even before we integrate over frequency and space.  Why is the divergence much worse in the theory of macro--QED than empty space QED?
	
\noindent\hrulefill\\

\subsubsection{Commutation relations\index{canonical commutation relations}}
The final loose end is to justify our assumption that the commutation relation between \(\hat{C}_{\omega}\) and \(\hat{C}_{\omega}^{\dagger}\) is given by the bosonic commutation relation (\ref{sarh:Ccom2}).  For this assumption to be consistent, the commutation relations for the field and reservoir variables (\ref{sarh:com1}--\ref{sarh:com2}) must not be altered when the field and reservoir operators are written in terms of \(\hat{C}_{\omega}\) and \(\hat{C}_{\omega}^{\dagger}\).  Taking the vector potential and its canonical momentum, and using their representation in terms of the creation and annihilation operators, (\ref{sarh:eopsol}) and (\ref{sarh:pezopsol}), one obtains
\begin{equation}
	\left[\hat{A}_{z}(x,t),\hat{\Pi}_{A_{z}}(x',t)\right]=\frac{i}{\pi c}\int_{0}^{\infty}d\omega \left|N_{\omega}\right|^{2}\omega\sqrt{\varepsilon(\omega)}e^{i\frac{\omega}{c}\sqrt{\varepsilon(\omega)}|x-x'|}-\text{c.c.}\label{sarh:commutation}
\end{equation}
where `\(-\,\text{c.c.}\)' implies the subtraction of the complex conjugate, and to obtain this formula we applied the result,
\begin{equation}
	\int_{-\infty}^{\infty}dx_{1}e^{i\frac{\omega}{c}\left(\sqrt{\varepsilon(\omega)}|x-x_{1}|-\sqrt{\varepsilon^{\star}(\omega)}|x'-x_{1}|\right)}=\frac{\sqrt{\varepsilon^{\star}(\omega)}e^{i\frac{\omega}{c}\sqrt{\varepsilon(\omega)}|x-x'|}}{\frac{\omega}{c}\text{Im}[\varepsilon(\omega)]}+\text{c.c.}\label{sarh:identity1}
\end{equation}
which using the notation of (\ref{sarh:1dgreen}) is equivalent to
\begin{equation}
	\int_{-\infty}^{\infty}dx_{1}\text{Im}[\varepsilon(\omega)]g(x-x_{1},\omega)g^{\star}(x'-x_{1},\omega)=\frac{c^{2}}{\omega^{2}}\text{Im}[g(x-x',\omega)].\label{sarh:green-identity}
\end{equation}
In passing we note that in the above form (\ref{sarh:green-identity}) is a result that can be generalised to two and three dimensions, as well as to inhomogeneous media, provided that the Green function vanishes at infinity.  To make progress we choose our undetermined constant \(N_{\omega}\) to take the value,
\begin{equation}
	N_{\omega}=\sqrt{\frac{\hbar}{2\omega}}\label{sarh:normalisation},
\end{equation}
a choice that is partly motivated by the fact that it allows us to write the right hand side of equation (\ref{sarh:commutation}) as an integral over the entire real line,
\begin{equation}
	\left[\hat{A}_{z}(x,t),\hat{\Pi}_{A_{z}}(x',t)\right]=\frac{i\hbar}{2\pi c}\int_{-\infty}^{\infty}d\omega \sqrt{\varepsilon(\omega)}e^{i\frac{\omega}{c}\sqrt{\varepsilon(\omega)}|x-x'|}.\label{sarh:integral-real-line}
\end{equation}
%

%
%
\begin{figure}[h]
	\begin{center}
	\includegraphics[width=8cm]{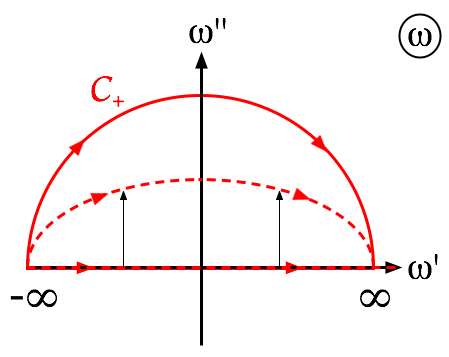}
	\caption{The integral along the real line given by (\ref{sarh:integral-real-line}) can be deformed into a contour integral, \(\mathcal{C}_{+}\) running along an infinite semicircle in the upper half frequency plane, \(\omega=\omega^{'}+i\omega^{''}\), by virtue of the fact that the integrand is analytic in this region.\label{sarh:contourfig}}
	\end{center}
\end{figure}

Assuming that \(\varepsilon(\omega)\) is free from zeros and poles in the upper half frequency plane~\footnote{This is true when \(\varepsilon(\omega)\) has a positive imaginary part, see~\cite{sarh:volume5} and exercise~\ref{sarh:epsexercise}.}, the integrand is analytic in this region, and the path of integration can be deformed into a semicircular contour, \(\mathcal{C}_{+}\), running from \(-\infty\) to \(\infty\) through the upper half plane\footnote{See e.g.~\cite{sarh:dettman1984}.}, as shown in Figure~\ref{sarh:contourfig}. Having done this we can see that the commutation relation gives the desired result,
\begin{align}
	\left[\hat{A}_{z}(x,t),\hat{\Pi}_{A_{z}}(x',t)\right]&=\frac{\hbar}{2\pi c}\int_{0}^{\pi}d\theta e^{i\theta}\lim_{|\omega|\to\infty}\left[|\omega| \sqrt{\varepsilon(\omega)}e^{i\frac{|\omega|e^{i\theta}}{c}\sqrt{\varepsilon(\omega)}|x-x'|}\right]\nonumber\\
	&=\frac{i\hbar}{\pi}\delta(x-x')\int_{0}^{\pi}d\theta\nonumber\\
	&=i\hbar\delta(x-x')\label{sarh:commutation-result}.
\end{align}
To get to the second line of (\ref{sarh:commutation-result}) we used (\ref{sarh:eps}) to show that \(\varepsilon(|\omega|\to\infty)=1\) in the upper half frequency plane, and applied the following representation of the delta function,
\[
	\delta(x)=\lim_{\lambda\to\infty}\frac{\lambda}{2}e^{-\lambda |x|}.
\]
It is worth noting that the integral identity (\ref{sarh:identity1}) (upon which (\ref{sarh:commutation-result}) rests) is not valid at frequencies where \(\text{Im}[\varepsilon(\omega)]=0\).  The validity of this theory therefore depends upon there being \emph{some} dissipation at all frequencies~\footnote{With the exception of \(\omega=0\) where there can be no dissipation.}.  The reader is left to verify that the commutation relations for the reservoir operators also do not change from (\ref{sarh:com2}).  The transformed Hamiltonian (\ref{sarh:diagham}) is therefore equivalent to the original expression (\ref{sarh:qH}), and we have now completed the first step towards our goal: we have shown how to develop a quantum mechanical theory of light in materials that includes the effects of dispersion and dissipation~\footnote{Further details on various aspects of this theory, along with its applications can be found in the extensive review paper of Scheel and Buhmann~\cite{sarh:scheel2008}.}.

\noindent\hrulefill
\paragraph*{Exercise:}  Verify that when written as (\ref{sarh:xopsol}) and (\ref{sarh:pxopsol}) the reservoir operators continue to satisfy (\ref{sarh:com2}).\\
	\emph{Warning:} This may take some time.\\
	\emph{Hint}: After evaluating the integrals over the delta functions, you'll still be left with integrals over frequency --- combine them and use the residue theorem. 
	
\noindent\hrulefill\\
%
%

\subsubsection{The ground state of the system\label{sarh:apsec}}

Before graduating from the safety of our 1D theory, we shall apply it to the ground state of the field and medium (the `vacuum state'), the properties of which are responsible for Casimir forces.  Our Hamiltonian (\ref{sarh:diagham}) is that of a continuum of uncoupled simple harmonic oscillators, and the ground state of this system (denoted by \(|0\rangle\)) can therefore be defined via
\begin{equation}
	\hat{C}_{\omega}(x)|0\rangle=0,\label{sarh:ground-state}
\end{equation}
i.e. if we try to find a state with less current in the medium we get identically zero.  Due to Heisenberg's uncertainty principle, the quantum mechanical ground state of a simple harmonic oscillator has an irreducible spread of possible positions and momenta.  In this case the material is represented by a continuum of simple harmonic oscillators, the ground state of which exhibits an irreducible fluctuation in the current.  The electromagnetic field (which can be thought of as originating from this current) therefore also has a spread of possible values, the properties of which we now calculate.

In accordance with their definitions in terms of the vector potential, the electric and magnetic field operators are given by\index{electric field operator}\index{magnetic field operator}
\begin{equation}
	\hat{E}_{z}=-\frac{\partial\hat{A}_{z}}{\partial t},\hspace{1cm}\hat{B}_{y}=-\frac{\partial\hat{A}_{z}}{\partial x}.\label{sarh:field-op}
\end{equation}
For this particular case we'll concentrate on the properties of the electric field
\begin{equation}
	\hat{E}_{z}(x,t)=-\mu_{0}\int_{0}^{\infty}d\omega\,\omega^{2} \sqrt{\frac{\hbar}{2\omega}}\alpha(\omega)\int_{-\infty}^{\infty}dx'g(x-x',\omega)\hat{C}_{\omega}(x')e^{-i\omega t}+\text{h.c.}\label{sarh:Ezop}
\end{equation}
the expectation value of which is zero in the ground state,
\begin{equation}
	\langle0|\hat{E}_{z}(x,t)|0\rangle=0.
\end{equation}
Meaning that although the field itself is not zero, if we did many measurements at a fixed point inside the medium we would find the average value to be zero.

Yet if we perform two measurements of the field at separate points \(x_{1}\) and \(x_{2}\), multiply the results together and then average we find a non--zero \emph{correlation}:
\begin{align}
	\langle0|\hat{E}_{z}(x_{1},t)\hat{E}_{z}(x_{2},t)|0\rangle&=\frac{\mu_{0}}{c^{2}}\int_{0}^{\infty}d\omega\,\omega^{4}\frac{\hbar}{\pi}\text{Im}[\varepsilon(\omega)]\int_{-\infty}^{\infty}dx'g(x_{1}-x',\omega)g^{\star}(x_{2}-x',\omega)\nonumber\\
	&=\frac{\hbar\mu_{0}}{\pi}\int_{0}^{\infty}d\omega\omega^{2}\text{Im}[g(x_{1}-x_{2},\omega)]\label{sarh:ecorr},
\end{align}
where we have applied our earlier result (\ref{sarh:identity1}) to obtain the second line.  The ground state of the field is thus spatially \emph{correlated} within the medium, with the value of the field at one point in space being related to the value at another point.  The integrand of (\ref{sarh:ecorr}) equals the result that would have been obtained from an application of the so--called linear fluctuation--dissipation theorem~\cite{sarh:volume5}, and represents the correlation of the field at a fixed frequency.  As an aside, notice that (\ref{sarh:ecorr}) can be continuously brought towards the limit \(\varepsilon(\omega)\to1\), so that the theory also applies to empty space\footnote{This limit is delicate.  As a rule of thumb, never take the limit \(\text{Im}[\varepsilon(\omega)]\to0\) in macro--QED until the end of a calculation.}.

As \(x_{1}\) and \(x_{2}\) approach one another, this correlation becomes a measure of the electric field intensity,
\begin{equation}
	\lim_{x_{1}\to x_{2}}\langle0|\hat{E}_{z}(x_{1})\hat{E}_{z}(x_{2})|0\rangle=\frac{\hbar\mu_{0}c}{2\pi}\int_{0}^{\infty}d\omega\,\omega\,\text{Im}\left[\frac{i}{\sqrt{\varepsilon(\omega)}}\right]\label{sarh:intensity},
\end{equation}
which diverges because there is a contribution to the field intensity at every frequency, which does not fall to zero as \(\omega\) increases.  This intensity has an associated field energy, which is also infinite, and is part of the divergent ground state energy in Eq. (\ref{sarh:diagham}).  This divergent contribution must be dealt with in calculations of the Casimir effect via a formal procedure called \emph{renormalisation}, which we shall return to in section~\ref{sarh:vacuum-force}.

\begin{samepage}
\noindent\hrulefill
\paragraph*{Exercise:}Show that the ground state expectation value of the following operator
	\[
		\hat{S}_{x}(x,t)=-\frac{1}{2\mu_{0}}\left[\hat{E}_{z}(x,t)\hat{B}_{y}(x,t)+\hat{B}_{y}(x,t)\hat{E}_{z}(x,t)\right]
	\]
	is zero.  What is the physical interpretation for this?
	
\noindent\hrulefill\\
\end{samepage}

As we have already established, the source of this field can be effectively thought of as a current within the medium.  The operator for this current can be inferred from the right hand side of the equation of motion for \(\hat{A}_{z}\) (\ref{sarh:eopeqm}), and is
\[
	\hat{j}(x,t)=-i\int_{0}^{\infty}d\omega\,\alpha(\omega)\sqrt{\frac{\hbar\omega}{2}}\left[\hat{C}_{\omega}(x)e^{-i\omega t}-\hat{C}^{\dagger}_{\omega}(x)e^{i\omega t}\right].
\]
This quantity also has zero expectation value,
\[
	\langle0|\hat{j}(x,t)|0\rangle=0,
\]
but its correlation function is proportional to a delta function, meaning that it is \emph{not} correlated in space
\begin{equation}
	\langle0|\hat{j}(x_{1},t)\hat{j}(x_{2},t)|0\rangle=\delta(x_{1}-x_{2})\frac{\hbar\varepsilon_{0}}{\pi}\int_{0}^{\infty}d\omega\,\omega^{2}\text{Im}[\varepsilon(\omega)].
\end{equation}
The reason for the lack of spatial correlation is that the excitations of the medium are independent from one another~\footnote{Had we dropped this assumption we would have introduced some \emph{spatial dispersion} into the material properties, which is a dependence of the permittivity on wave--vector as well as field, and amounts to an extra spatial correlation in the field within the medium~\cite{sarh:volume8,sarh:horsley2014}.}.  Meanwhile the electromagnetic field obeys a wave equation containing spatial derivatives and as a consequence different points are not independent, leading to the correlation (\ref{sarh:ecorr}).
%
%
\subsection{Macroscopic QED in three dimensions\label{sarh:fullsec}}

The extension of the results given in section \ref{sarh:1dsec} to three dimensional electromagnetism is fairly straightforward, for the price of little more than an occasionally cumbersome notation. In three dimensions the Lagrangian for the field is of the same overall form as (\ref{sarh:field}), but with the electric and magnetic fields being given in terms of both the scalar\index{scalar potential} and vector\label{sarh:Magnetic vector potential} potentials
\begin{equation}
	\vec{E}=-\vec{\nabla}\varphi-\dot{\vec{A}}\hspace{2cm}\vec{B}=\vec{\nabla}\vec{\times}\vec{A}.
\end{equation}
For our purposes we can think of the potentials as a useful shorthand for the fields, although in general we must emphasise that in quantum mechanics the potentials are the more fundamental quantities.  The three parts of the Lagrangian density (\ref{sarh:ld}) are now given by
\begin{align}
	\mathscr{L}_{F}&=\frac{\varepsilon_{0}}{2}\left[\left(\vec{\nabla}\varphi+\dot{\vec{A}}\right)^{2}-c^{2}(\vec{\nabla}\vec{\times}\vec{A})^{2}\right]\label{sarh:lf}\\
	\mathscr{L}_{R}&=\frac{1}{2}\int_{0}^{\infty}\left[\left(\frac{\partial\vec{X}_{\omega}}{\partial t}\right)^{2}-\omega^{2}\vec{X}_{\omega}^{2}\right]d\omega\label{sarh:lr}\\
	\mathscr{L}_{I}&=-(\vec{\nabla}\varphi+\dot{\vec{A}})\vec{\cdot}\int_{0}^{\infty}\alpha(\omega)\vec{X}_{\omega}d\omega\label{sarh:li}
\end{align}
where \(\varphi\), \(\vec{A}\), and \(\vec{X}_{\omega}\) are all functions of position and time.  Nothing is very different: the effect of the material on the field is again mimicked with a reservoir but the amplitudes of the simple harmonic oscillators are now vectors, and the expression for the field now includes the scalar potential.  In general a second reservoir should be added to account for losses through the magnetic permeability~\footnote{See e.g.~\cite{sarh:kheirandish2008,sarh:philbin2010}.} \(\mu(\omega)\), although here we have neglected the magnetic properties of the medium.  

Applying the Euler--Lagrange equations to (\ref{sarh:lf}--\ref{sarh:li}), we find the equations of motion for the electromagnetic field are given by,
\begin{equation}	\vec{\nabla}\vec{\cdot}\left[\varepsilon_{0}\vec{E}+\int_{0}^{\infty}\alpha(\omega)\vec{X}_{\omega}d\omega\right]=0\label{sarh:divD}
\end{equation}
and
\begin{equation}
	\vec{\nabla}\vec{\times}\vec{B}-\frac{1}{c^{2}}\frac{\partial\vec{E}}{\partial t}=\mu_{0}\frac{\partial}{\partial t}\int_{0}^{\infty}d\omega\,\alpha(\omega)\vec{X}_{\omega}\label{sarh:field-eqns}.
\end{equation}
The remaining two Maxwell equations listed in (\ref{sarh:macmax}) are identically true when the fields are written in terms of the potentials.  The equation for the reservoir is the vector generalisation of the one dimensional case, and again has the solution (\ref{sarh:oscgreen}--\ref{sarh:Xsol})
\begin{equation}
	\vec{X}_{\omega}(\vec{x},t)=\frac{\alpha(\omega)}{\omega}\int_{0}^{\infty}\sin(\omega\tau)\vec{E}(\vec{x},t-\tau)d\tau+\vec{C}_{\omega}(\vec{x})e^{-i\omega t}+\vec{C}^{\star}_{\omega}(\vec{x})e^{i\omega t}\label{sarh:Xvsol}.
\end{equation}
Substituting this expression for \(\vec{X}_{\omega}\) into the electromagnetic field equations (\ref{sarh:divD}--\ref{sarh:field-eqns}) gives us the behaviour of the electromagnetic field in the medium without reference to the reservoir.  These are the macroscopic Maxwell equations,
\begin{align}	
	\vec{\nabla}\vec{\cdot}\vec{D}&=\rho_{f}\nonumber\\
	\vec{\nabla}\vec{\times}\vec{H}&=\vec{j}_{f}+\frac{\partial\vec{D}}{\partial t}\label{sarh:f12}.
\end{align}
The \(\vec{H}\) field is defined as simply proportional to the magnetic field \(\vec{H}=\vec{B}/\mu_{0}\), and the \emph{displacement field}\index{displacement field} \(\vec{D}\) as
\begin{equation}
	\vec{D}(\vec{x},t)=\varepsilon_{0}\left[\vec{E}(\vec{x},t)+\int_{0}^{\infty}d\tau\chi(\tau)\vec{E}(\vec{x},t-\tau)\right]\label{sarh:DH}.
\end{equation}
The quantity \(\chi(\tau)\) is equal to (\ref{sarh:rsus}) so that the coupling function between the reservoir and the field \(\alpha(\omega)\) is given by \(([2\omega\varepsilon_{0}\text{Im}[\varepsilon(\omega)])/\pi]^{1/2}\) which is the same expression we derived in one dimension (\ref{sarh:coupling}).  The free charge and current density, \(\rho_{f}\) and \(\vec{j}_{f}\) in (\ref{sarh:f12}) --- i.e. the current and charge density \emph{not} induced by the field --- are equal to
\begin{align}
	\rho_{f}(\boldsymbol{x},t)&=-\vec{\nabla}\vec{\cdot}\int_{0}^{\infty}d\omega\,\alpha(\omega)\vec{C}_{\omega}(\boldsymbol{x})e^{-i\omega t}+\text{c.c.}\nonumber\\
	\vec{j}_{f}(\boldsymbol{x},t)&=-i\int_{0}^{\infty}d\omega\,\omega\alpha(\omega)\vec{C}_{\omega}(\boldsymbol{x})e^{-i\omega t}+\text{c.c.}\label{sarh:rhoj}
\end{align}
and automatically satisfy the continuity equation, \(\vec{\nabla}\vec{\cdot}\vec{j}+\partial\rho/\partial t=0\).  The reservoir amplitudes \(\vec{C}_{\omega}\) have the same interpretation as before, which is now explicit in equation (\ref{sarh:rhoj}): they make up the amplitude of the free electric current density within the medium, which is responsible for the electromagnetic field.  For notational brevity we have not indicated any spatial dependence for \(\alpha(\omega)\), but all the results given in this section continue to hold when this is a function of position as it is for inhomogeneous media.

To reiterate the point made earlier, remember that we are assuming some degree of dissipation at all frequencies and all points in space, so any field must come from a source. In our case the system is just field plus material, so the source can only be some oscillating current within the material, which is given by \(\vec{j}_{f}\).  In macro--QED we can thus think of the Casimir effect as being dictated by the interaction of the fluctuating currents within the two semi--infinite plates, across the gap between them. It is worth contrasting this picture from that which is ordinarily used to understand the Casimir effect, where the plates simply serve to restrict the allowed modes of the field. Our modification to the traditional understanding is a necessary consequence of properly including dispersion and dissipation.

The derivation of the Hamiltonian from (\ref{sarh:lf}--\ref{sarh:li}) produces the same result as (\ref{sarh:1dham}) and the same canonical momenta as (\ref{sarh:Emom}--\ref{sarh:Xmom}), but again with scalar quantities becoming vectors.  In three dimensions the quantum mechanical Hamiltonian operator therefore takes the same form as (\ref{sarh:qH}),
\begin{multline}
	\hat{H}=\frac{1}{2}\int d^{3}\vec{x}\bigg\{\frac{1}{\varepsilon_{0}}\left(\hat{\vec{\Pi}}_{\vec{A}}+\int_{0}^{\infty}d\omega\,\alpha(\omega)\hat{\vec{X}}_{\omega}\right)^{2}+\frac{1}{\mu_{0}}\left(\vec{\nabla}\vec{\times}\hat{\vec{A}}\right)^{2}\\
	+\int_{0}^{\infty}d\omega\left[\hat{\vec{\Pi}}_{\vec{X}_{\omega}}^{2}+\omega^{2}\hat{\vec{X}}_{\omega}^{2}\right]\bigg\}.\label{sarh:3dmacham}
\end{multline}

\subsubsection{Gauge condition\index{gauge condition}}

Everything is now essentially a matter of listing slightly generalised versions of the operator formulae given in section~\ref{sarh:1dsec} --- except for a slight niggle, which may have already occurred to the reader: \emph{what happened to the scalar potential in the Hamiltonian?}  It's disappeared!  This is not an oddity confined to macro--QED~\footnote{See e.g.~\cite{sarh:ryder2003}.}.  The reason for its absence is that the Lagrangian does not contain \(\dot{\varphi}\), so the associated canonical momentum is identically zero,
\begin{equation}
	\hat{\Pi}_{\varphi}=0.\label{sarh:phicanmom}
\end{equation}
Equation (\ref{sarh:phicanmom}) implies that the `equation of motion' associated with \(\hat{\varphi}\), \(\vec{\nabla}\vec{\cdot}\hat{\vec{D}}=\hat{\rho}_{f}\), is not to be understood an equation of motion at all, but must be interpreted as the relationship between \(\hat{\varphi}\) and \(\hat{\vec{A}}\), that allows us to eliminate the scalar potential \(\hat{\varphi}\) from the Hamiltonian.  In order to make sense of this condition on the potentials we fix a gauge, \(\vec{\nabla}\vec{\cdot}\hat{\vec{A}}=0\)~\footnote{This is often called the \emph{Coulomb gauge}.} so that the equation for the divergence of \(\hat{\vec{D}}\) implies that the scalar potential can be elimated and written in terms of the reservoir,
\begin{equation}
	\vec{\nabla}\hat{\varphi}=\frac{1}{\varepsilon_{0}}\left(\int_{0}^{\infty}\alpha(\omega)\hat{\vec{X}}_{\omega}d\omega\right)_{L},\label{sarh:constraint}
\end{equation}
where a subscript `\(L\)' indicates the longitudinal part of the vector~\footnote{The longitudinal part of a vector field \(\vec{V}\) is that part which has divergence, but no curl.  In terms of a Fourier expansion of the function this is
\[\vec{V}_{L}(\vec{x})=\int\frac{d^{3}\vec{k}}{(2\pi)^{3}}\frac{1}{k^{2}}\vec{k}[\vec{k}\vec{\cdot}\tilde{\vec{V}}(\vec{k})]e^{i\vec{k}\vec{\cdot}\vec{x}}e^{i\vec{k}\vec{\cdot}\vec{x}},\]where \(\tilde{\vec{V}}\) is the Fourier amplitude of \(\vec{V}\).}.  The expression for the scalar potential given by (\ref{sarh:constraint}) was imposed to obtain the $\hat{\varphi}$ independent Hamiltonian, (\ref{sarh:3dmacham}).  

\begin{samepage}
\noindent\hrulefill
\paragraph*{Exercise:} Using condition (\ref{sarh:constraint}), derive the Hamiltonian (\ref{sarh:3dmacham}) from the Lagrangian given by the sum of (\ref{sarh:lf}--\ref{sarh:li}).

\noindent\hrulefill\\
\end{samepage}

\subsubsection{Commutation relations}

Having fixed a gauge to eliminate the scalar potential from the Hamiltonian, we must also make sure the commutation relations between \(\hat{\vec{A}}\) and \(\hat{\vec{\Pi}}_{\vec{A}}\) are consistent with this gauge, i.e. \([\vec{\nabla}\vec{\cdot}\hat{\vec{A}},\hat{\vec{\Pi}}_{\vec{A}}]=0\).  A consistent set of commutation relations is given by\footnote{See for instance~\cite{sarh:ryder2003,sarh:weinberg1995,sarh:leonhardt2010}.}
\begin{equation}
	\left[\hat{\vec{A}}(\vec{x},t),\hat{\vec{\Pi}}_{\vec{A}}(\vec{x}',t)\right] = i\hbar\,\tens{\delta}_{T}(\vec{x}-\vec{x}'),\label{sarh:3dcomm}
\end{equation}
where \(\tens{\delta}_{T}(\vec{x}-\vec{x}')\) is the \emph{transverse delta function}~\footnote{The transverse part of a vector field is that part which has curl but zero divergence.}, which is a rank two tensor.  There is no such complication with the reservoir operators, which are not constrained in this way, and obey the expected generalisation of (\ref{sarh:com2}),
\begin{equation}
	\left[\hat{\vec{X}}_{\omega}(\vec{x},t),\hat{\vec{\Pi}}_{\vec{X}_{\omega'}}(\vec{x}',t)\right]=i\hbar\tens{1}_3\delta^{(3)}(\vec{x}-\vec{x}')\delta(\omega-\omega').
\end{equation}
where $\tens{1}_3$ is a $3\times3$ identity matrix.

\begin{samepage}
\noindent\hrulefill
\paragraph*{Exercise:} Starting from the following ansatz
	\[
		\left[\hat{\vec{A}}(\vec{x},t),\hat{\vec{\Pi}}(\vec{x}',t)\right]=i\hbar\left[\tens{1}_{3}\delta^{(3)}(\vec{x}-\vec{x}')-\boldsymbol{M}(\vec{x}-\vec{x}')\right]
	\]
	show that the two constraints \(\vec{\nabla}\vec{\cdot}\hat{\vec{A}}=0\) and \(\vec{\nabla}\vec{\cdot}\hat{\vec{D}}=0\) imply
	\begin{align*}
		\vec{k}&=\vec{k}\vec{\cdot}\tilde{\boldsymbol{M}}(\vec{k})\\
		\vec{k}&=\vec{k}\vec{\cdot}\tilde{\boldsymbol{M}}^{T}(\vec{k})
	\end{align*}
	where
	\[
		\boldsymbol{M}(\vec{x}-\vec{x}')=\int\frac{d^{3}\vec{k}}{(2\pi)^{3}}\tilde{\boldsymbol{M}}(\vec{k})e^{i\vec{k}\vec{\cdot}(\vec{x}-\vec{x}')}.
	\]
	Assuming that both indices of \(\tilde{\boldsymbol{M}}\) are longitudinal, then show that (\ref{sarh:3dcomm}) must be the correct commutation relation.
	
\noindent\hrulefill\\
\end{samepage}

\subsubsection{Diagonalising the Hamiltonian\index{diagonalising the Hamiltonian}}

In the quantum mechanical case, the electromagnetic field and reservoir operators still obey the classical equations of motion (see problem~\ref{sarh:quadprob}).  Their expressions in terms of the creation and annihilation operators for excitations in the reservoir can be found through taking the solutions to the macroscopic Maxwell equations (\ref{sarh:f12}) in terms of \(\vec{C}_{\omega}\), and replacing these amplitudes with \(\sqrt{\hbar/2\omega}\) times the operator \(\hat{\vec{C}}_{\omega}\), just as we did in section~\ref{sarh:1dsec}.  The operators that result from this process are
\begin{align}
	\hat{\vec{E}}(\vec{x},t)&=i\mu_{0}\int_{0}^{\infty}d\omega\,\omega\int d^{3}\vec{x}'\tens{G}(\vec{x},\vec{x}',\omega)\vec{\cdot}\hat{\vec{j}}_{f}(\vec{x'},\omega)e^{-i\omega t}+\text{h.c.}\nonumber\\
	\hat{\vec{B}}(\vec{x},t)&=\mu_{0}\int_{0}^{\infty}d\omega\int d^{3}\vec{x}'\vec{\nabla}\vec{\times}\tens{G}(\vec{x},\vec{x}',\omega)\vec{\cdot}\hat{\vec{j}}_{f}(\vec{x'},\omega)e^{-i\omega t}+\text{h.c.}\label{sarh:3dfop}
\end{align}
and
\begin{equation}
	\hat{\vec{X}}_{\omega}(\vec{x},t)=\frac{\alpha(\omega)}{\omega}\int_{-\infty}^{t}\sin[\omega(t-t')]\hat{\vec{E}}(\vec{x},t')dt'+\sqrt{\frac{\hbar}{2\omega}}\left[\hat{\vec{C}}_{\omega}(\vec{x})e^{-i\omega t}+\hat{\vec{C}}_{\omega}^{\dagger}(\vec{x})e^{i\omega t}\right]\label{sarh:3drop}.
\end{equation}
So again the theory works in terms of quanta of current within the medium, from which the field is determined.  The electromagnetic Green function \(\tens{G}(\vec{x},\vec{x}',\omega)\) is a rank two object with two vector indices, obeying (in a non--magnetic material)
\begin{equation}
	\vec{\nabla}\times\vec{\nabla}\times\tens{G}-k_0^2\varepsilon\tens{G}=\tens{1}\delta^{(3)}(\vec{x}-\vec{x}')
\end{equation}
The operator corresponding to the Fourier amplitude of the free electrical current \(\hat{\vec{j}}_{f}\) is defined as
\begin{equation}
	\hat{\vec{j}}_{f}(\vec{x},\omega)=-i\sqrt{\frac{\hbar\omega}{2}}\alpha(\omega)\hat{\vec{C}}_{\omega}(\vec{x}).\label{sarh:jop}
\end{equation}
The form of the Hamiltonian in terms of the creation and annihilation operators can again be inferred from the time--dependence of the operators (\ref{sarh:3dfop}--\ref{sarh:3drop}) and is the same as (\ref{sarh:diagham}),
\begin{equation}
	\hat{H}=\frac{1}{2}\int d^{3}\vec{x}\int_{0}^{\infty}d\omega\,\hbar\omega\,\hat{\vec{C}}_{\omega}(\vec{x})\vec{\cdot}\hat{\vec{C}}_{\omega}^{\dagger}(\vec{x})+\text{h.c.}\label{sarh:3dh}
\end{equation}
which can be justified in exactly the same way as (\ref{sarh:diagham}), with \(\hat{\vec{C}_{\omega}}\) and \(\hat{\vec{C}}_{\omega}^{\dagger}\) satisfying
\begin{equation}
	\left[\hat{\vec{C}}_{\omega}(\vec{x},t),\hat{\vec{C}}_{\omega'}^{\dagger}(\vec{x}',t)\right]={\tensunit}\delta^{(3)}(\vec{x}-\vec{x}')\delta(\omega-\omega')\label{sarh:3dCC}.
\end{equation}
Although this account of the full theory is somewhat cursory, we trust the reader can understand its meaning on the basis of what went before it.  The above formulae encompass the theory of light in absorbing media, a theory which we have developed from a Hamiltonian that self--consistently includes the effects of dissipation and dispersion.  This theory provides an underlying theoretical framework for the quantum versions of the macroscopic Maxwell equations (\ref{sarh:macmax}), and one that may be extended to unambiguously treat moving objects.  The application to moving objects, and the forces between them is the purpose of the second half of this chapter.

\begin{samepage}
\noindent\hrulefill
\paragraph*{Exercise:}Starting from the expression for the electric field operator given by (\ref{sarh:3dfop}), show that the correlation of the electric field in the ground state \(\langle0|\hat{\vec{E}}(\vec{x},t)\tprod\hat{\vec{E}}(\vec{x}',t)|0\rangle\) is that predicted by the fluctuation--dissipation theorem~\cite{sarh:volume5},
	\[
		\langle0|\hat{\vec{E}}(\vec{x},t)\tprod\hat{\vec{E}}(\vec{x}',t)|0\rangle=\frac{\hbar\mu_{0}}{\pi}\int_{0}^{\infty}d\omega\,\omega^{2}\text{Im}\left[\tens{G}(\vec{x},\vec{x}',\omega)\right].
	\]\\
	\emph{Hint}: Use the integral identity for Green functions (\ref{sarh:green-identity2}).\\

\paragraph*{Exercise:}
	Starting from the following general expression for a quadratic Hamiltonian,
	\[
		\hat{H}=\sum_{i,j}\left\{\alpha_{ij}\hat{p}_{i}\hat{p}_{j}+\beta_{ij}\hat{q}_{i}\hat{q}_{j}+\frac{1}{2}\gamma_{ij}\left[\hat{p}_{i}\hat{q}_{j}+\hat{q}_{j}\hat{p}_{i}\right]\right\}+\sum_{i}\left[V_{i}\hat{q}_{i}+W_{i}\hat{p}_{i}\right]
	\]
	where \(\alpha_{ij}\), \(\beta_{ij}\) and \(\gamma_{ij}\) are arbitrary symmetric constant matrices, and \(V_{i}\) and \(W_{i}\) are constant vectors, show that the operators obey the classical equations of motion.\label{sarh:quadprob}
	
\noindent\hrulefill\\
\end{samepage}
%
%
%
%
{\footnotesize
\noindent
\textbf{Example --- a single polariton:}  Given our rather brief synopsis of macro--QED in three dimensions, an example might be helpful.  We could calculate some ground state property of the system, but we won't learn anything fundamentally new compared to our one dimensional theory.
\par
Consider the simplest possible excitation of the system above the ground state: a single excitation of the medium, with the current aligned along the \(\vec{e}_{z}\) axis,
\begin{equation}
	|\psi\rangle=\int_{0}^{\infty} d\omega\int d^{3}\vec{x} f(\vec{x},\omega)\vec{e}_{z}\vec{\cdot}\hat{\vec{C}}_{\omega}^{\dagger}(\vec{x})|0\rangle\label{sarh:state1},
\end{equation}
where \(f(\vec{x},\omega)\) is a function that is sharply peaked around \(\vec{x}_{0}\) and \(\omega_{0}\).  To ensure that \(\langle\psi|\psi\rangle=1\), the function \(f(\vec{x},\omega)\) must be normalised to one,
\begin{equation}
	\int d^{3}\vec{x}\int_{0}^{\infty}d\omega|f(\vec{x},\omega)|^{2}=1.\label{sarh:norm}
\end{equation}
This is also the lowest level of excitation of the field. We'll call this excitation a \emph{polariton}\index{polariton}, which is the name coined by Hopfield~\cite{sarh:hopfield1958} for a mixture of electromagnetic and material excitation.  The properties of the field in this state can be gleaned from the electric field correlation function which, after a few steps, we find to be
\begin{multline}
	\langle\psi|\hat{\vec{E}}(\vec{x}_{1})\vec{\otimes}\hat{\vec{E}}(\vec{x}_{2})|\psi\rangle=\frac{\hbar\mu_{0}}{\pi}\bigg[\frac{1}{2}\int_{0}^{\infty}d\omega\,\omega^{2}\text{Im}[\tens{G}(\vec{x}_{1},\vec{x}_{2},\omega)]\\
	+\int_{0}^{\infty}d\omega\int_{0}^{\infty}d\omega'\int d^{3}\vec{x}'\int d^{3}\vec{x}''\frac{\omega^{2}\omega'^{2}}{c^{2}}\sqrt{\text{Im}[\varepsilon(\omega)]\text{Im}[\varepsilon(\omega')]}e^{i(\omega'-\omega)t}\\
	\times f^{\star}(\vec{x}',\omega)f^{\star}(\vec{x}'',\omega')\tens{G}(\vec{x}_{1},\vec{x}',\omega)\vec{\cdot}\vec{e}_{z}\vec{\otimes}\vec{e}_{z}\vec{\cdot}\tens{G}^{\dagger}(\vec{x}_{2},\vec{x}'',\omega')\bigg]+\text{c.c.}\label{sarh:epolcor}
\end{multline}
where we applied the three dimensional generalisation of result (\ref{sarh:green-identity}),
\begin{equation}
	\frac{\omega^{2}}{c^{2}}\int d^{3}\vec{x}'\text{Im}[\varepsilon(\omega)]\tens{G}(\vec{x}_{1},\vec{x}',\omega)\vec{\cdot}\tens{G}^{\dagger}(\vec{x}_{2},\vec{x}',\omega)=\text{Im}[\tens{G}(\vec{x}_{1},\vec{x}_{2},\omega)]\label{sarh:green-identity2}.
\end{equation}
To proceed we write the expansion coefficient \(f\) as a product of Gaussians,
\begin{equation}
	f(\vec{x},\omega)=\left(\frac{1}{\pi(\Delta x)^{2}}\right)^{3/4}\left(\frac{2}{\pi(\Delta\omega)^{2}}\right)^{1/4}e^{-\frac{1}{2(\Delta x)^{2}}(\vec{x}-\vec{x}_{0})^{2}}e^{-\frac{1}{2(\Delta\omega)^{2}}(\omega-\omega_{0})^{2}}\label{sarh:f-func}
\end{equation}
and assume that \(\Delta x\) and \(\Delta\omega\) are small enough that all the functions in (\ref{sarh:epolcor}) are constant over the region where \(f\) is significantly different from zero.  Carrying out the integrations in (\ref{sarh:epolcor}) we obtain the final expression for the correlation function,
\begin{equation}
	\langle\psi|\hat{\vec{E}}(\vec{x}_{1})\vec{\otimes}\hat{\vec{E}}(\vec{x}_{2})|\psi\rangle=\frac{\hbar\mu_{0}}{\pi}\bigg[\int_{0}^{\infty}d\omega\,\omega^{2}\text{Im}[\tens{G}(\vec{x}_{1},\vec{x}_{2},\omega)]+\mathcal{B}\,\text{Re}\left[\tens{G}(\vec{x}_{1},\vec{x}_{0},\omega_{0})\vec{\cdot}\vec{e}_{z}\vec{\otimes}\vec{e}_{z}\vec{\cdot}\tens{G}^{\dagger}(\vec{x}_{2},\vec{x}_{0},\omega_{0})\right]\bigg]\label{sarh:epolcor2}
\end{equation}
where \(\mathcal{B}=32\sqrt{2}\pi^{2}\Delta x^{3}\,\Delta\omega\,\omega_{0}^{4}\,\text{Im}[\varepsilon(\omega_{0})]/c^{2}\).  There are several interesting things about (\ref{sarh:epolcor2}).  Firstly, the limit of a point--like excitation \(\Delta x\to0\), or one of infinitesimal bandwidth, \(\Delta\omega\to0\) just gives back the vacuum correlation function.  This result has its roots in the normalisation of the state (\ref{sarh:state1}), and means that the quantum states of light within an absorbing medium must have a finite bandwidth, and originate from a source of non--zero extent.  Secondly, notice that the correlation function breaks up into a sum of a vacuum contribution (the three dimensional version of (\ref{sarh:intensity})) plus an additional term arising from the excitation in the medium.  Therefore the correlation of the field is a superposition of the vacuum correlation plus that of the polariton.

While the total intensity of the field diverges in the limit \(\vec{x}_{1}\to\vec{x}_{2}=\vec{x}\), the difference in intensity between the ground and excited states is finite,
\begin{equation}
	\langle\psi|\hat{\vec{E}}(\vec{x})\vec{\otimes}\hat{\vec{E}}(\vec{x})|\psi\rangle-\langle0|\hat{\vec{E}}(\vec{x})\vec{\otimes}\hat{\vec{E}}(\vec{x})|0\rangle=\frac{\hbar\mu_{0}\mathcal{B}}{\pi}\,\text{Re}\left[\tens{G}(\vec{x},\vec{x}_{0},\omega_{0})\vec{\cdot}\vec{e}_{z}\vec{\otimes}\vec{e}_{z}\vec{\cdot}\tens{G}^{\dagger}(\vec{x},\vec{x}_{0},\omega_{0})\right]\label{sarh:s1c},
\end{equation}
and is exactly what one would obtain for the time average of the classical electric field intensity from a current distributed in space and frequency according to (\ref{sarh:f-func}), with an amplitude proportional to, \((\hbar\,\text{Im}[\varepsilon(\omega)])^{1/2}\). (see problem~\ref{sarh:sourceprob})}

\begin{samepage}
\noindent\hrulefill
	\paragraph*{Exercise:}  Consider an electric field due to a current \(\vec{j}(\vec{x},\omega)=\vec{e}_{z}\mathcal{J}(\vec{x},\omega)\) where \(\mathcal{J}(\vec{x},\omega)\) is sharply peaked around \(\vec{x}_{0}\) and \(\omega_{0}\).  Show that the electric field is of the form
	\[
		\vec{E}(\vec{x},t)=A\tens{G}(\vec{x},\vec{x}_{0},\omega_{0})\vec{\cdot}\vec{e}_{z}e^{-i\omega_{0}t}+\text{c.c.}
	\]
	where \(A\) is proportional to the amplitude of the current.  From this expression show that the time average of \(\vec{E}\vec{\otimes}\vec{E}\) is
	\[
		\langle\vec{E}(\vec{x})\vec{\otimes}\vec{E}(\vec{x})\rangle=|A|^{2}\tens{G}(\vec{x},\vec{x}_{0},\omega_{0})\vec{\cdot}\vec{e}_{z}\vec{\otimes}\vec{e}_{z}\vec{\cdot}\tens{G}^{\dagger}(\vec{x},\vec{x}_{0},\omega_{0})+\text{c.c.}
	\]\label{sarh:sourceprob}
which shows that the average electric field intensity in this state is the same as for a classical point source at position $\vec{x}_0$.

\noindent\hrulefill\\
\end{samepage}
%
%
\section{Vacuum forces between moving bodies}\label{sarh:vacforcemovbod}
\par
The relative motion of macroscopic bodies implies a non--equilibrium situation~\footnote{See e.g.~\cite{sarh:volume5}.} and if we were to apply e.g. the fluctuation--dissipation theorem~\cite{sarh:volume5}, which is derived for systems in thermal equilibrium, it would in general have to be with care.  In typical calculations of the Casimir force it is imagined that the bodies experiencing the force are held at rest, and we calculate the external force required to maintain this situation.  Therefore in such calculations, all the usual equilibrium results apply.  However, macro--QED is not restricted to the equilibrium state and can in--principle treat quantum forces between objects in relative motion.  We finish the tutorial with a calculation of the quantum force between two bodies in relative motion.

Armed with the theory of the electromagnetic field in realistic media, we now find what effect the quantum field has on the motion of a body.  We shall derive a quantum theory of electromagnetic forces through modifying the Lagrangian used in the previous section.
%
%
\subsection{Moving bodies in 1D macroscopic QED\label{sarh:mbmqed}}
\par
Returning to electromagnetism in one dimension, imagine that the homogeneous medium that we previously investigated is set into motion with uniform velocity \(V\) along the \(x\) axis.  The \emph{value} of the Lagrangian density will be the same (it is a scalar under Lorentz transformations), but it will \emph{look} different when written in this new reference frame.  This modification can be found through rewriting the earlier Lagrangian (\ref{sarh:field}--\ref{sarh:coupling1}) in a relativistically covariant form\footnote{For a recap of relativistic notation\index{covariant form} see~\cite{sarh:volume2}.}
\begin{align}
	\mathscr{L}_{F}&=\frac{1}{2\mu_{0}}(\partial_{\mu}A_{z})(\partial^{\mu}A_{z})\label{sarh:lfm}\\
	\mathscr{L}_{I}&=-V^{\mu}(\partial_{\mu}A_{z})\int_{0}^{\infty}\alpha(\omega)X_{\omega}d\omega\label{sarh:lim}\\
	\mathscr{L}_{R}&=\frac{1}{2}\int_{0}^{\infty}\left[\left(V^{\mu}\partial_{\mu}X_{\omega}\right)^{2}-\omega^{2}X_{\omega}^{2}\right]d\omega,\label{sarh:lrm}
\end{align}
where we have defined \(V^{\mu}=\gamma(c,V)\), \(\partial_{\mu}=(c^{-1}\partial_{t},\partial_{x})\), \(\partial^{\mu}=(c^{-1}\partial_{t},-\partial_{x})\), \(\gamma=(1-V^{2}/c^{2})^{-1/2}\), and a repeated Greek index implies summation (the Einstein summation convention\index{Einstein summation convention}).  The coupling between the medium and the field takes the rest frame value \(\alpha(\omega)=[2\omega\varepsilon_{0}\text{Im}[\varepsilon'(\omega)]/\pi]^{1/2}\) where \(\varepsilon'(\omega)\) is the permittivity measured in the rest frame, and the quantities \(V^{\mu}\) and \(\partial_{\mu}\) are \emph{four--vectors}\index{four--vector}, although in this 1D case only the `\(x\)' and `\(t\)' components are important.  When \(V=0\) then (\ref{sarh:lfm}--\ref{sarh:lrm}) reduce to the Lagrangian density given by (\ref{sarh:field}--\ref{sarh:coupling1}).

\begin{samepage}
\noindent\hrulefill
\paragraph*{Exercise:} The Lagrangian is often constructed through taking the difference between the kinetic and potential energy of a system.  Consider a particle of mass \(m\) at rest.  If we associate the energy \(mc^{2}\) with the particle the action in the rest frame must be \(S=-\int mc^{2}dt'\) (the minus sign is a matter of convention, and \(t'\) is the time in the rest frame).  Show that when written in covariant form this is
	\[
		S=-\int m V_{\mu}dx^{\mu}=-\int mc^{2}\sqrt{1-\vec{V}^{2}/c^{2}}dt
	\]
	which is the relativistic action for a free particle.  The equations of motion for the particle can be found through varying this action with respect to \(\vec{V}\).  Now use a similar argument to derive (\ref{sarh:lfm}--\ref{sarh:lrm}) from their rest frame counterparts.\\

\paragraph*{Exercise:} The effect of the motion of the medium appears in (\ref{sarh:lfm}--\ref{sarh:lrm}) as the operator \(V^{\mu}\partial_{\mu}\).  What is the physical meaning of this operator?  

\noindent\hrulefill\\
\end{samepage}

The motion of the medium influences the electromagnetic field in two ways: (i) the motion of the reservoir is modified, because absorbed energy is now carried along with the medium rather than remaining at a fixed point; and (ii) the coupling to the reservoir is modified, because a moving dielectric medium polarises in response to both electric and magnetic fields.  These modifications are evident in the equations of motion for the field derived from the above Lagrangian,
\begin{equation}
	\left[\frac{\partial^{2}}{\partial x^{2}}-\frac{1}{c^{2}}\frac{\partial^{2}}{\partial t^{2}}\right]A_{z}=-\mu_{0}\left(\frac{\partial}{\partial t}+V\frac{\partial}{\partial x}\right)\int_{0}^{\infty}\alpha(\omega)X_{\omega}d\omega\label{sarh:mveqm1}
\end{equation}
and the reservoir
\begin{equation}
	\left[\left(\frac{\partial}{\partial t}+V\frac{\partial}{\partial x}\right)^{2}+\omega^{2}\right]X_{\omega}=-\alpha(\omega)\left(\frac{\partial}{\partial t}+V\frac{\partial}{\partial x}\right) A_{z}\label{sarh:mveqm2}
\end{equation}
where we have assumed the velocity of the medium is slow enough such that \(\gamma\sim1\).  The general solution to (\ref{sarh:mveqm2}) can be written as
\begin{multline}
	X_{\omega}(x,t)=-\alpha(\omega)\int_{-\infty}^{\infty} dx_{0}\int_{0}^{\infty} d \tau\mathcal{G}_{R}(x-x_{0},\tau)\left(\frac{\partial}{\partial t}+V\frac{\partial}{\partial x_{0}}\right)A_{z}(x_{0},t-\tau)\\[5pt]
	+C_{\omega}(x-Vt)e^{-i\omega t}+C^{\star}_{\omega}(x-Vt)e^{i\omega t}\label{sarh:Xmsol}
\end{multline}
where \(\mathcal{G}_{R}\) is the retarded Green function of equation (\ref{sarh:mveqm2}),
\begin{equation}
	\mathcal{G}_{R}(x-x_{0},\tau)=\Theta(\tau)\frac{\sin(\omega\tau)}{\omega}\delta\left(x-x_{0}-V\tau\right).\label{sarh:gfoscmm}
\end{equation}
As before, \(\mathcal{G}_{R}\) is the response of \(X_{\omega}\) (which mimics the medium) to a sudden force applied at the time \(\tau=0\) at the point \(x=x_{0}\).  The argument of the delta function is such that the excitation of any current is carried along with the material, remaining at e.g. a single point, but moving with velocity \(V\).

\begin{samepage}
\noindent\hrulefill
\paragraph*{Exercise:} Show that (\ref{sarh:gfoscmm}) satisfies
	\[
		\left[\left(\frac{\partial}{\partial t}+V\frac{\partial}{\partial x}\right)^{2}+\omega^{2}\right]\mathcal{G}_{R}(x-x_{0},t-t_{0})=\delta(t-t_{0})\delta(x-x_{0}).
	\]
\noindent\hrulefill
\end{samepage}

Substituting the reservoir motion (\ref{sarh:Xmsol}) into the wave equation for the electromagnetic vector potential (\ref{sarh:mveqm1}) we obtain the equation for the field, allowing us to identify the susceptibility of the moving medium
\begin{equation}
	\left[\frac{\partial^{2}}{\partial x^{2}}-\frac{1}{c^{2}}\frac{\partial^{2}}{\partial t^{2}}\right]A_{z}=\frac{1}{c^{2}}\int_{-\infty}^{\infty} dx_{0}\int_{0}^{\infty} d\tau\chi(x-x_{0},\tau)\left(\frac{\partial}{\partial t}+V\frac{\partial}{\partial x_{0}}\right)^{2}A_{z}(x_{0},t-\tau)\label{sarh:wveq2},
\end{equation}
where \(C_{\omega} =0\) and
\begin{equation}
	\chi(x-x_{0},\tau)=\frac{2}{\pi}\int_{0}^{\infty}d\omega\text{Im}[\varepsilon'(\omega)]\sin(\omega\tau)\delta\left(x-x_{0}-V\tau\right).
\end{equation}
Next to the susceptibility in (\ref{sarh:wveq2}) we have both spatial and temporal derivatives of the vector potential, which is because a moving material polarises in response to both electric and magnetic fields~\cite{sarh:volume8}.  From this susceptibility we can identify the effective permittivity as we did in (\ref{sarh:epsr}), which is now a function of both \(\omega\) and \(k\),
\begin{align}
	\varepsilon(k,\omega)&=1+\int_{-\infty}^{\infty} dx \int_{0}^{\infty}d\tau\chi(x-x_{0},\tau)e^{i\omega\tau}e^{-ik(x-x_{0})}\nonumber\\
	&=1+i\,\text{Im}[\varepsilon'(\omega-Vk)]+\frac{2}{\pi}\text{P}\int_{0}^{\infty}\frac{\Omega\text{Im}[\varepsilon'(\Omega)]}{\Omega^{2}-(\omega-Vk)^{2}}d\Omega.\label{sarh:epsmoveq}
\end{align}
This is \emph{exactly} the same as our earlier expression in a non--moving material (\ref{sarh:epsr}) but with the frequency shifted from \(\omega\) to \(\omega-Vk\).  Our Lagrangian thus describes the physical phenomenon where the dispersion of a moving material is \emph{Doppler shifted} relative to the rest frame.  Indeed, we shall find that the same terms in the Lagrangian that are responsible for this Doppler shift in frequency can be used to predict the \emph{force} on a moving body!  The next section will illustrate that in general the Doppler effect is inseparable from the physics of radiation pressure.  Note that although we have concentrated our efforts on the case of constant velocity the above approach can be equally well applied to any arbitrary motion of the body, so long as it remains much less than \(c\).
%
%
\subsection{Computing classical forces}\label{sarh:comclafor}
%
%
\begin{figure}[h]
	\begin{center}
	\includegraphics[width=8cm]{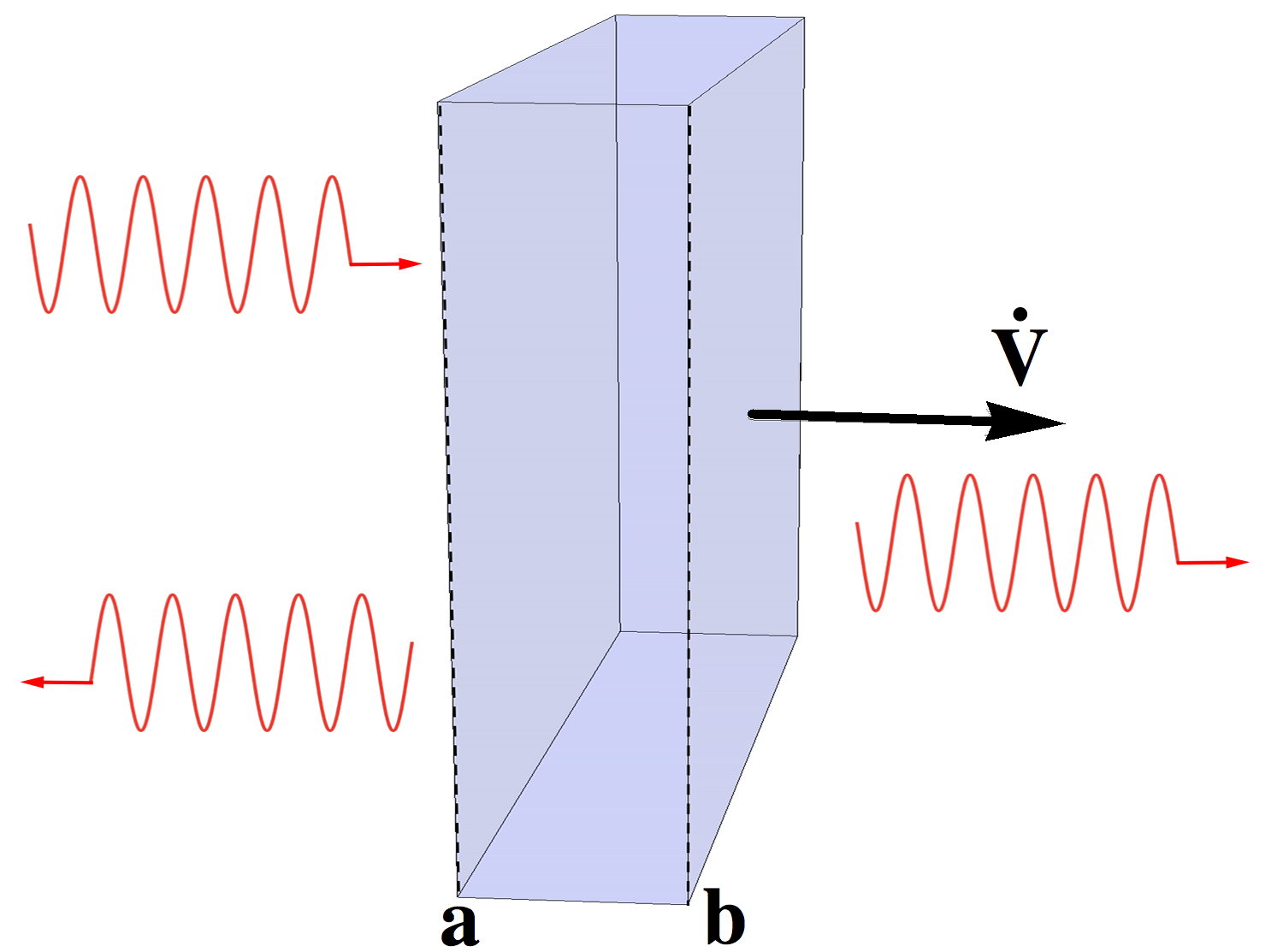}
	\caption{Applying the Lagrangian of macro--QED we find that the force on the centre of mass, \(M\dot{V}\), is the Lorentz force integrated over the volume of the body.  This force can be re--written in terms of the difference in the electromagnetic stress, \(\sigma_{xx}\) on the two sides, \(\sigma_{xx}(b)-\sigma_{xx}(a)\), minus the rate of change of the electromagnetic momentum within the body.  In section~\ref{sarh:qrp} we show that the operator equivalent of the Lorentz force is the appropriate expression for the quantum mechanical case.\label{sarh:forcefig}}
	\end{center}
\end{figure}
\par
Through making a slight modification to the theory of a uniformly moving medium, we can also describe electromagnetic forces.  To do this we simply take the Lagrangian for a uniformly moving homogeneous medium (\ref{sarh:lfm}--\ref{sarh:lrm}), generalise it to the case of an inhomogeneous medium \(\alpha(\omega)\to\alpha(\omega,x-R(t))\) (\(R\) is the centre of mass coordinate) and add the centre of mass kinetic energy
\begin{equation}
	L=\frac{1}{2}MV^{2}+\int dx\left[\mathscr{L}_{F}+\mathscr{L}_{I}+\mathscr{L}_{R}\right]\label{sarh:1dml}
\end{equation}
where in this one dimensional case, \(M\) is the mass per unit cross--sectional area of the body.  Varying the new Lagrangian with respect to \(R\) and \(V\) then gives us the equation of motion for the position of the body
\[
	\frac{d}{d t}\left(\frac{\partial L}{\partial V}\right)=\frac{\partial L}{\partial R}
\]
which, from our earlier Lagrangian density (\ref{sarh:lrm}), yields an expression that initially looks quite complicated
\begin{multline}
	M\dot{V}+\frac{d}{dt}\int dx\int_{0}^{\infty}d\omega\left[\frac{\partial X_{\omega}}{\partial x}\left(\frac{\partial X_{\omega}}{\partial t}+V\frac{\partial X_{\omega}}{\partial x}\right)-\frac{\partial A_{z}}{\partial x}\alpha(\omega,x-R)X_{\omega}\right]\\[5pt]
	=-\int dx\left(\frac{\partial A_{z}}{\partial t}+V\frac{\partial A_{z}}{\partial x}\right)\int_{0}^{\infty}\frac{\partial\alpha(\omega,x-R)}{\partial R}X_{\omega}d\omega.\label{sarh:1deqm1}
\end{multline}
However, applying the equation of motion for the reservoir (\ref{sarh:mveqm2}) and enforcing the boundary condition \(X_{\omega}=\Pi_{X_{\omega}}=0\) at spatial infinity allows us to simplify this down to
\begin{align}
	M\dot{V}&=\int d x\frac{\partial A_{z}}{\partial x}\int_{0}^{\infty}d\omega\,\alpha(\omega,x-R)\left(\frac{\partial X_{\omega}}{\partial t}+V\frac{\partial X_{\omega}}{\partial x}\right)\nonumber\\
	&=-\int d x j_{z}B_{y}\label{sarh:force1}.
\end{align}
This is simply the total \emph{Lorentz force}~\footnote{In general the Lorentz force density is given by \(\vec{f}_{L}=\rho\vec{E}+\vec{j}\times\vec{B}\) which reduces to \(f_{L}=-j_{z}B_{y}\) in this 1D case.}!  The current density appearing in this expression for the force is
\begin{equation}
	j_{z}=\int_{0}^{\infty}\alpha(\omega,x-R)\left(\frac{\partial X_{\omega}}{\partial t}+V\frac{\partial X_{\omega}}{\partial x}\right)\label{sarh:currmov}
\end{equation}
which is the rest frame current already identified in (\ref{sarh:currres}), but written in the coordinate system where the medium is in motion (problem~\ref{sarh:jtransprob}).

\begin{samepage}
\noindent\hrulefill
\paragraph*{Exercise:} Verify that (\ref{sarh:force1}) follows from (\ref{sarh:1deqm1}).\\

\paragraph*{Exercise:} Using the fact that \(\rho=0\) in this 1D case, show that the Lorentz transformation of the current (\ref{sarh:currres}) from the rest frame to one where the material is in motion gives (\ref{sarh:currmov}) when \(\gamma\sim1\).\\

\emph{Hint}: The current transforms between frames as \(j_{z}'=\gamma(j_{z}-V\rho)\).\label{sarh:jtransprob}

\noindent\hrulefill\\
\end{samepage}

Recall that the velocity dependence occurs within the Lagrangian in both the coupling between field and reservoir, and in the dynamics of the reservoir itself.  It is the latter of these that is the origin of the modified frequency dependence of the permittivity (\ref{sarh:epsmoveq}).  \emph{From this we can draw the conclusion that the force on a dielectric body is fundamentally linked to the Doppler effect}.

\subsubsection{The stress tensor and the Poynting vector}

Using the equation for the vector potential (\ref{sarh:mveqm1}), we can replace the current \(j_{z}\) in (\ref{sarh:force1}) with \(-\mu_{0}^{-1}[\partial_{x}^{2}-c^{-2}\partial_{t}^{2}]A_{z}\), and the electromagnetic force on the material (\ref{sarh:force1}) can then be re--written in terms of the fields alone,
\begin{align}
	M\dot{V}&=\frac{1}{\mu_{0}}\int d x\left(\frac{1}{c^{2}}\frac{\partial E_{z}}{\partial t}-\frac{\partial B_{y}}{\partial x}\right)B_{y}\nonumber\\
	&=\int dx\frac{\partial T_{xx}}{\partial x}-\frac{1}{c^{2}}\frac{\partial}{\partial t}\int S_{x}dx\label{sarh:stf},
\end{align}
where the electromagnetic \emph{stress}\index{stress tensor} is identified as
\begin{equation}
	T_{xx}=-\frac{\varepsilon_{0}}{2}\left[E_{z}^{2}+c^{2}B_{y}^{2}\right],\label{sarh:st1}
\end{equation}
which is a single component of the three dimensional \emph{stress tensor}\footnote{See, e.g.~\cite{sarh:volume2}.} \(\tens{T}\), and the electric and magnetic fields are defined as in (\ref{sarh:field-def}).  Meanwhile the \(x\)--component of the \emph{Poynting vector}\index{Poynting vector} is
\begin{equation}
	S_{x}=-\frac{1}{\mu_{0}}E_{z}B_{y},
\end{equation}
which represents the electromagnetic power per unit area flowing in the \(x\) direction.  The two integrands on the right hand side of (\ref{sarh:stf}) identically cancel in the region of space where there is no medium, as can be verified from the wave equation in the absence of a source.  Therefore the integrals may be taken over the body alone, which in this case we assume to extend from \(x=a\) to \(b\) (as in figure~\ref{sarh:forcefig}) giving
\begin{equation}
	M\dot{V}=T_{xx}(b)-T_{xx}(a)-\dot{\mathcal{P}}\label{sarh:stf2},
\end{equation}
where
\[
	\mathcal{P}=\frac{1}{c^{2}}\int_{a}^{b} S_{x}dx.
\]
We can interpret the force given in (\ref{sarh:stf2}) as the difference in radiation pressure on the two sides of the body minus the rate of change of the net electromagnetic momentum \(\mathcal{P}\) within the body~\footnote{Being concerned with mechanical forces we identify the momentum density in the medium with the Abraham expression \(\vec{E}\vec{\times}\vec{H}/c^{2}\).  For further details on the momentum of light in media see~\cite{sarh:barnett2010}.}.  After time averaging --- which removes \(\mathcal{P}\) from the equation --- this is the result one would obtain from the classical theory of radiation pressure~\footnote{See e.g.~\cite{sarh:volume8,sarh:novotny2006}.}, but here we have derived it from an action principle that is set up to self--consistently describe the effects of material dispersion and dissipation.  Having constructed the Lagrangian of macro--QED, we got a theory of forces for free!  There was no need to postulate a form for the stress tensor; this came automatically from our description of moving media.  We can also quantise this theory, thus obtaining a quantum theory of radiation pressure that is not restricted to any particular state of the field or motion of the body, which is a distinct advantage of applying macro--QED to calculate quantum forces.

\begin{samepage}
\noindent\hrulefill
\paragraph*{Exercise:}  Consider a wave incident onto a material occupying the space \(0<x<a\)
	\[
		A_{z}=\frac{E_{0}}{2i\omega}\begin{cases}\left[e^{i\frac{\omega}{c}x}+re^{-i\frac{\omega}{c}x}\right]e^{-i\omega t}+\text{c.c.}&x<0\\
				te^{i\frac{\omega}{c}x}e^{-i\omega t}+\text{c.c.}&x>a
				\end{cases}
	\]
	where \(r\) and \(t\) are the reflection and transmission coefficients of the interface.  Show that for such a field the time average of the stress (\ref{sarh:st1}) is given by
	\[
		\langle T_{xx}\rangle=-\frac{\varepsilon_{0}}{2}\left|E_{0}\right|^{2}\begin{cases}
						1+|r|^{2}&x<0\\
						|t|^{2}&x>a
						\end{cases}
	\]
	and therefore that the force per unit area imparted by this field is proportional to \(1+|r|^{2}-|t|^{2}\).  Can you give an interpretation for this result?
	
\noindent\hrulefill\\
\end{samepage}
%
%
\subsection{Quantum theory of radiation pressure\label{sarh:qrp}}
\par
To quantise this theory we take the same approach as in section~\ref{sarh:1dsec}, and first construct the Hamiltonian.

\subsubsection{Classical Hamiltonian}

The canonical momenta of the field and the reservoir are modified by the motion of the medium
\begin{align*}
	\Pi_{A_{z}}&=\frac{\partial\mathscr{L}}{\partial\dot{A}_{z}}=\varepsilon_{0}\frac{\partial A_{z}}{\partial t}-\int_{0}^{\infty}\alpha(\omega,x-R)X_{\omega}d\omega\\
	\Pi_{X_{\omega}}&=\frac{\partial\mathscr{L}}{\partial\dot{X}_{\omega}}=\frac{\partial X_{\omega}}{\partial t}+V\frac{\partial X_{\omega}}{\partial x}
\end{align*}
and the momentum associated with the centre of mass is
\begin{equation}
	p=\frac{\partial L}{\partial V}=MV+\mathcal{A},
\end{equation}
where
\begin{equation}
	\mathcal{A}=\int d x\int_{0}^{\infty}d\omega\left[\frac{\partial X_{\omega}}{\partial x}\Pi_{X_{\omega}}-\frac{\partial A_{z}}{\partial x}\alpha(\omega,x-R)X_{\omega}\right]\label{sarh:Gamma}.
\end{equation}
Applying these expressions within the definition of the Hamiltonian, we find
\begin{align}
	H&=p V+\int dx\left(\Pi_{A_{z}}\dot{A}_{z}+\int_{0}^{\infty}d\omega\dot{X}_{\omega}\Pi_{X_{\omega}}\right)-L\nonumber\\
	&=\frac{(p-\mathcal{A})^{2}}{2M}+\int\mathscr{H}_{0}dx\label{sarh:Vham},
\end{align}
where \(\mathscr{H}_{0}\) is given by the earlier expression for a stationary medium (\ref{sarh:1dham}).

Interestingly the above Hamiltonian (\ref{sarh:Vham}) is of the same \emph{form} as that of a charged particle in an electromagnetic field~\cite{sarh:volume2}, but it describes the centre of mass motion of a macroscopic body\index{Hamiltonian, moving medium}.  In this respect the quantity \(\mathcal{A}\) is analogous to the vector potential.  For a charged particle the vector potential can be thought of as the momentum carried by the charge due to its interaction with the field~\cite{sarh:semon1996}. Analogously \(\mathcal{A}\) is the part of the momentum carried by the centre of mass due to its coupling to both field and material degrees of freedom.  In the Hamiltonian formalism the motion of the medium is coupled to the field and the reservoir (the internal degrees of freedom of the material) through the quantity \(\mathcal{A}\).  As an aside it is worth noting that the term in \(\mathcal{A}\) that equals \(\partial_{x} A_{z}\alpha(\omega)X_{\omega}\) is the macroscopic version of the R\"ontgen interaction that occurs between a single moving electric dipole and a magnetic field~\cite{sarh:wilkens1994}\index{R\"ontgen interaction}.  

\subsubsection{Hamiltonian operator}

The Hamiltonian operator can be inferred from its classical counterpart (\ref{sarh:Vham}) and takes the form
\begin{equation}
	\hat{H}=\frac{(\hat{p}-\hat{\mathcal{A}})^{2}}{2M}+\hat{H}_{0}\label{sarh:qHV},
\end{equation}
where
\begin{equation}
	\hat{\mathcal{A}}=\int d x\int_{0}^{\infty}d\omega\left[\frac{1}{2}\left(\hat{\Pi}_{X_{\omega}}\frac{\partial \hat{X}_{\omega}}{\partial x}+\frac{\partial \hat{X}_{\omega}}{\partial x}\hat{\Pi}_{X_{\omega}}\right)-\frac{\partial \hat{A}_{z}}{\partial x}\alpha(\omega,x-\hat{R})\hat{X}_{\omega}\right]\label{sarh:qGV}
\end{equation}
and \(\hat{H}_{0}\) is given by the expression for a stationary body (\ref{sarh:qH}).  The only difference in the form of (\ref{sarh:qHV}--\ref{sarh:qGV}) compared to the classical case is that we have chosen a symmetric ordering of \(\hat{X}_{\omega}\) and \(\hat{\Pi}_{X_{\omega}}\) in \(\hat{\mathcal{A}}\).  For completeness, we note that the commutation relation between the centre of mass \(\hat{R}\) and canonical momentum \(\hat{p}\) takes the usual value
\begin{equation}
	\left[\hat{R},\hat{p}\right]=i\hbar\label{sarh:movingcm}.
\end{equation}

\subsubsection{Operator equations of motion}

It is now possible to apply the theory to the problem of quantum electromagnetic forces on objects.  The motion of the centre of mass can be determined from the equations of motion for the centre of mass operator, \(\hat{R}\), the expectation value of which gives us the average position of the body.

The Hamiltonian gives us both the velocity of the centre of mass
\begin{equation}
	\frac{d\hat{R}}{dt}=\frac{i}{\hbar}\left[\hat{H},\hat{R}\right]=\frac{1}{M}\left(\hat{p}-\hat{\mathcal{A}}\right)\label{sarh:vcm}
\end{equation}
and the acceleration
\begin{align}
	\frac{d^{2}\hat{R}}{dt^{2}}&=\frac{i}{M\hbar}\left[\hat{H},\hat{p}-\hat{\mathcal{A}}\right]\nonumber\\
	&=-\frac{1}{M}\int dx\int_{0}^{\infty}d\omega\frac{\partial\alpha(\omega,x-\hat{R})}{\partial R}\frac{\partial\hat{A}_{z}}{\partial t}\hat{X}_{\omega}-\frac{i}{M\hbar}\left[\hat{H}_{0},\hat{\mathcal{A}}\right].\label{sarh:accop1}
\end{align}
Evaluating the commutation relations and using the commutator identity \([\hat{A},\hat{B}\hat{C}]=[\hat{A},\hat{B}]\hat{C}+\hat{B}[\hat{A},\hat{C}]\), the acceleration of the body can be written as
\begin{align}
	\frac{d^{2}\hat{R}}{dt^{2}}&=\frac{1}{M}\int dx\int_{0}^{\infty}d\omega\,\alpha(\omega,x-\hat{R})\frac{\partial \hat{A}_{z}}{\partial x}\hat{\Pi}_{X_{\omega}}\nonumber\\
	&=-\frac{1}{M}\int dx\hat{j}_{z}\hat{B}_{y}\label{sarh:acm},
\end{align}
which is the operator equivalent of the classical force (\ref{sarh:force1}), where
\[
	\hat{j}_{z}=\int_{0}^{\infty}d\omega\,\alpha(\omega,x-\hat{R})\hat{\Pi}_{X_{\omega}}.
\]
This leads us to the conclusion that the force on the centre of mass of a moving body is determined by the Lorentz force operator.  This agrees --- for example --- with the work of Loudon~\cite{sarh:loudon2002} on quantum mechanical radiation pressure which assumes that the force is given by the expectation value of the Lorentz force operator\index{Lorentz force operator}.

\begin{samepage}
\noindent\hrulefill
\paragraph{Exercise:}  Fill in the steps between (\ref{sarh:accop1}) and (\ref{sarh:acm}).

\noindent\hrulefill\\
\end{samepage}

As in the classical case discussed above, we can write the acceleration of the body entirely in terms of the field.  Through applying the equation of motion for the vector potential operator
\[
	\frac{\partial^{2}\hat{A}_{z}}{\partial x^{2}}-\frac{1}{c^{2}}\frac{\partial\hat{A}_{z}}{\partial t^{2}}=-\mu_{0}\int_{0}^{\infty}\alpha(\omega,x-\hat{R})\hat{\Pi}_{X_{\omega}}
\]
we find an expression that is formally identical to the classical result,
\begin{equation}
	M\frac{d^{2}\hat{R}}{dt^{2}}=\hat{T}_{xx}(b)-\hat{T}_{xx}(a)-\frac{\partial\hat{\mathcal{P}}}{\partial t}\label{sarh:qrp1}
\end{equation}
where the energy momentum tensor operator is defined as
\begin{equation}
	\hat{T}_{xx}=-\frac{\varepsilon_{0}}{2}\left[\hat{E}_{z}^{2}+c^{2}\hat{B}_{y}^{2}\right]\label{sarh:qstress}
\end{equation}
and the integrate Poynting vector operator as
\[
	\hat{\mathcal{P}}=\frac{1}{c^{2}}\int_{a}^{b}\hat{S}_{x}dx
\]
with the Poynting vector
\begin{equation}
	\hat{S}_{x}=-\frac{1}{2\mu_{0}}\left[\hat{E}_{z}\hat{B}_{y}+\hat{B}_{y}\hat{E}_{z}\right].\label{sarh:qpoynt}
\end{equation}
The force on the centre of mass of a body is thus equal to the expectation value of the operator equivalent of the classical radiation pressure (\ref{sarh:stf}).  This result is true whatever the state of the system.  Indeed, one could imagine small objects containing many atoms prepared such that the centre of mass behaves quantum mechanically, and the operator nature of \(\hat{R}\) becomes important~\footnote{A similar situation has been considered previously, for the case of a perfect mirror interacting with a quantised field~\cite{sarh:law1995}, with both the position of the mirror and the field imagined to be in a quantum state.  The theory developed above is a generalisation of this earlier work to the case of arbitrarily shaped bodies, characterised in terms of a permittivity obeying the Kramers--Kronig relations.}.

%
%
\subsection{The vacuum force\label{sarh:vacuum-force}}

The above theory is now applied to the simplest case of interest: the force on a body initially localised at a point \(R=R_{0}\), with the field and medium in the ground state.  The initial wave function\index{wave function} of the total system is taken to be of the form,
\[
	|\psi\rangle=\left(\frac{1}{\pi(\Delta x)^{2}}\right)^{1/4}e^{-\frac{1}{2(\Delta x)^{2}}(R-R_{0})^{2}}|0_{R_{0}}\rangle,
\]
where it is assumed that the localisation of the centre of mass \(\Delta x\) is much smaller than any other length scale of interest (i.e. the relevant wavelengths of the field), and \(|0_{R_{0}}\rangle\) is defined as the state where \(\hat{C}_{\omega}(x,R_{0})|0_{R_{0}}\rangle=0\).  The dependence of the creation and annihilation operators on \(R_{0}\) is necessary because the states of the whole system are different when the body is at different positions.  From the previous section the average value for the force on the body is the expectation value of (\ref{sarh:qrp1}),
\begin{equation}
	M\frac{d^{2}\langle\hat{R}\rangle}{dt^{2}}=\langle\psi|\left[\hat{T}_{xx}(b)-\hat{T}_{xx}(a)-\frac{\partial\hat{\mathcal{P}}}{\partial t}\right]|\psi\rangle\label{sarh:expect}.
\end{equation}
Applying our earlier expressions for the field operators (\ref{sarh:eopsol}), the expectation value of the stress and the Poynting vector are found to be
\begin{align}
	\langle\psi|\hat{T}_{xx}|\psi\rangle&=-\frac{\varepsilon_{0}}{2}\lim_{x\to x'}\langle0_{R_{0}}|\left[\hat{E}_{z}(x)\hat{E}_{z}(x')+c^{2}\hat{B}_{y}(x)\hat{B}_{y}(x')\right]|0_{R_{0}}\rangle\nonumber\\
	&=-\frac{\hbar}{2\pi c^{2}}\lim_{x\to x'}\int_{0}^{\infty}d\omega\left\{\omega^{2}\text{Im}[g(x,x',\omega)]+c^{2}\frac{\partial^{2}}{\partial x\partial x'}\text{Im}[g(x,x',\omega)]\right\}\label{sarh:stressop}
\end{align}
and
\[
	\langle\psi|S_{x}|\psi\rangle=0.
\]
To obtain the second line of (\ref{sarh:stressop}) we applied the Green function identity (\ref{sarh:green-identity}) and the same reasoning that led to the ground state correlation function (\ref{sarh:ecorr}), taking the limit \(x\to x'\) in the correlation function to obtain the field intensity.  The expectation value for the force is given by the difference in the stress (\ref{sarh:stressop}) on the two sides of the body, which is the one dimensional version of the \emph{Lifshitz theory}~\cite{sarh:volume}, used to compute quantum forces between stationary bodies, restricted to the case of material bodies separated by vacuum.  We have derived this result from a quantum mechanical theory based on a Hamiltonian derived from a classical action, treating all variables as operators.  In this calculation we have not included the other bodies in the Hamiltonian, but doing so does not change (\ref{sarh:stressop}).  The above result is formally valid for any system of bodies, and one need only use a Green function that satisfies (\ref{sarh:1dgreeneq}) with \(\varepsilon(x,\omega)\) defining the configuration of the objects.

\begin{samepage}
\noindent\hrulefill
\paragraph{Exercise:}  Describe how the above calculation would differ if the centre of mass of the body was prepared in a quantum mechanical state.  Explain why this is usually not important.

\noindent\hrulefill\\
\end{samepage}

\subsubsection{Renormalisation\index{renormalisation}}

One obvious --- and very serious --- problem with using the stress (\ref{sarh:stressop}) to calculate the force is that as it stands it is meaningless.  As we already discovered in section~\ref{sarh:1dsec}, these integrals over the Green functions diverge and therefore the expectation value for the stress (\ref{sarh:stressop}) in the force is infinite!  However we know that the force on the body cannot be infinite so the \emph{difference} between the stress on the two sides of the body must be a finite number (if it isn't then we are in serious trouble).  To attempt to remove this infinite but spatially uniform contribution to the stress tensor, we subtract part of the Green function,
\begin{equation}
	g(x,x',\omega)\to g(x,x',\omega)-g_{0}(x-x',\omega)\equiv g^{(S)}(x,x',\omega)\label{sarh:regular},
\end{equation}
a process we are referring to as \emph{renormalisation}~\cite{sarh:volume9}.  The function, \(g_{0}\), is the Green function for a homogeneous medium with the permittivity \(\varepsilon\) at the point of interest, and \(g^{(S)}\) can be thought of as the `scattered' part of the Green function.  We need to be very careful when we do this, because we are changing the theory by hand --- in general \emph{not} a good idea!  However, this modification does not change the end result for the force, because in the limit \(x\to x'\), \(g_{0}\) does not depend on \(x\), and therefore (\ref{sarh:regular}) does not modify the expression for the radiation pressure, \(\langle[\hat{\sigma}_{xx}(b)-\hat{\sigma}_{xx}(a)]\rangle\).  Although it seems in most cases that (\ref{sarh:regular}) yields a finite value for the force, at present there is reason to suspect that it can fail in some situations~\footnote{See e.g.~\cite{sarh:simpson2013}}.  

Note that here we have focussed on obtaining a quantum theory of radiation pressure, and (\ref{sarh:regular}) is merely a trick to extract the finite quantity of interest from the divergent stress.  This formal trick is distinct from what is typically referred to as renormalisation in the quantum field theory literature, which is required when the quantity of interest turns out infinite.  In that case the divergent contribution has to be absorbed into one or more of the physical constants~\footnote{See e.g.~\cite{sarh:ryder2003}}.

\subsection{A simple case of quantum friction\label{sarh:qfsec}}

When the field and medium are in their ground state then, in addition to the attractive force between two separated objects, there exists a \emph{frictional} force that serves --- for planar media --- to bring any relative lateral motion to zero.  This phenomenon is known as \emph{quantum friction}~\footnote{For further details see~\cite{sarh:levitov1989,sarh:braginsky1991,sarh:pendry1997,sarh:volokitin2008}.}\index{quantum friction}, and we shall now consider a simple instance of this effect within our one dimensional theory.

Suppose that the medium is at rest, and the field is initially in the zero particle state \(|0\rangle\).  Another system is coupled to it, which moves with a fixed velocity \(V>0\), and has an internal degree of freedom \(\xi\) which is proportional to its dipole moment (e.g. this could be a moving atom).  We imagine that the internal degree of freedom of this object is also in its ground state, and the coupling is switched on at \(t=0\).
%
%
\begin{figure}[h]
	\begin{center}
	\includegraphics[width=9cm]{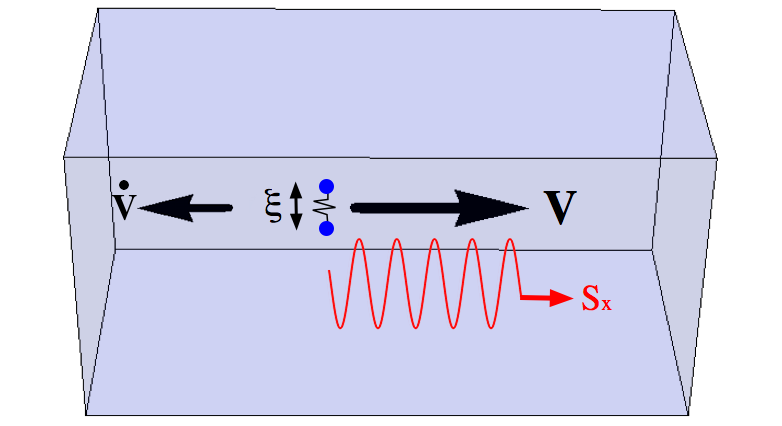}
	\caption{A polarisable particle is dragged through a material (e.g. water) at velocity V.  Even though both particle and field may initially be in their respective ground states, the motion of the particle leads to an emission of radiation in the same direction as the velocity, and the particle therefore experiences a frictional force \(\dot{v}\).\label{sarh:friction-fig}}
	\end{center}
\end{figure}

To describe this situation we add a new term into the Hamiltonian
\begin{equation}
	\hat{H}_{I}(t)=i\beta\Theta(t)(\hat{a}e^{-i\omega_{0}t}-\hat{a}^{\dagger}e^{i\omega_{0}t})\frac{\partial \hat{A}_{z}(Vt,t)}{\partial t}\label{sarh:Hint},
\end{equation}
where \(\beta\) is proportional to the particle's polarisability, and \(\hat{a}\) and \(\hat{a}^{\dagger}\) are the raising and lowering operators associated with the internal degree of freedom \(\xi\).  Planck's constant times the transition frequency \(\omega_{0}\) is the energy required to excite the internal degree of freedom.  This interaction Hamiltonian (\ref{sarh:Hint}) is written above in the \emph{interaction picture}~\footnote{See e.g.~\cite{sarh:volume4}.}, where the time dependence of all the operators is generated by the Hamiltonian without the interaction term.  Adopting a perturbative approach, to first order in \(\beta\) the state of the system after the interaction has been switched on is the ground state plus some superposition of excited states,
\begin{equation}
	|\psi\rangle=|0\rangle|0\rangle_{\xi}+\int \frac{d k}{2\pi}\int_{0}^{\infty}d\omega c(k,\omega,t)\hat{C}^{\dagger}_{\omega}(k)\hat{a}^{\dagger}|0\rangle|0\rangle_{\xi},\label{sarh:state}
\end{equation}
where \(|0\rangle_{\xi}\) is the ground state of the oscillator, defined as that state which the lowering operator reduces to zero \(\hat{a}|0\rangle_{\xi}=0\).  We have adopted a Fourier transformed representation in which the \(\hat{C}_{\omega}\) operators have been written as a function of \(k\) rather than \(x\) obeying
\[
	\left[\hat{C}_{\omega}(k),\hat{C}^{\dagger}_{\omega'}(k')\right]=2\pi\delta(k-k')\delta(\omega-\omega').
\]
The vector potential operator (\ref{sarh:eopsol}) becomes
\begin{equation}
	\hat{A}_{z}(x,t)=-i\mu_{0}\int_{0}^{\infty}d\omega\sqrt{\frac{\hbar\omega}{2}}\alpha(\omega)\int\frac{dk}{2\pi}g(k,\omega)\hat{C}_{\omega}(k)e^{ikx}e^{-i\omega t}+\text{h.c.}\label{sarh:Ak}
\end{equation}
where \(g(k,\omega)\) is the Fourier transform of (\ref{sarh:1dgreen})
\[
	g(k,\omega)=\frac{1}{k^{2}-\frac{\omega^{2}}{c^{2}}\varepsilon(\omega)}.
\]
In this representation the excitation of the system is encoded in the expansion coefficient \(c(k,\omega,t)\) which is given by the matrix element of the interaction Hamiltonian\footnote{See e.g.~\cite{sarh:volume4}.},
\begin{align}
	c(k,\omega,t)&=-\frac{i}{\hbar}\int_{-\infty}^{t}\langle0|_{\xi}\langle0|\hat{C}_{\omega}(k)\hat{a}\hat{H}_{I}(t')|0\rangle|0\rangle_{\xi}d t'\nonumber\\
	&=-i\mu_{0}\beta\sqrt{\frac{\omega}{2\hbar}}\alpha(\omega)(\omega-Vk)\Theta(t)\frac{e^{i(\omega_{0}+\omega-Vk)t}-1}{(\omega_{0}+\omega-Vk)}g^{\star}(k,\omega).
\end{align}
Already we can see that when \(V\neq0\), the transition amplitude, \(c(k,\omega,t)\) is peaked around the point where the Doppler shifted frequencies of the reservoir equal \emph{minus} the transition frequency of the particle \(\omega_{0}+\omega-Vk=0\).  This is the point where a positive frequency, \(\omega>0\), in the rest frame of the medium has been Doppler shifted to a negative value in the rest frame of the particle, \(\omega-Vk<0\).  To put it another way, a positive energy excitation of the field and medium, \(\hbar\omega\), appears as a negative energy excitation \(\hbar(\omega-Vk)\) in the rest frame of the polarisable object, which represents energy \emph{available} to excite the internal degree of freedom of the particle \(\xi\)Q  After a long time compared to the inverse of the transition frequency, the rate of this transition \(\Gamma_{0\to1}(k,\omega)\) is
\begin{align}
	\Gamma_{0\to1}&=\lim_{t\to\infty}\int_{0}^{\infty}d\omega\int_{-\infty}^{\infty}\frac{dk}{2\pi}\frac{|c(k,\omega,t)|^{2}}{t}\nonumber\\
	&\to\frac{2\mu_{0}\beta^{2}\omega_{0}^{2}}{\hbar}\int_{\omega_{0}/V}^{\infty}\frac{dk}{2\pi}\text{Im}[g(k,V k-\omega_{0})]\label{sarh:transition-rate},
\end{align}
which is zero when \(V=0\).  Note that to obtain (\ref{sarh:transition-rate}) we used the following representation of the delta function
\[
	\delta(x)=\frac{1}{\pi}\lim_{\lambda\to\infty}\frac{\sin^{2}(\lambda x)}{\lambda x^{2}}.
\]
The result (\ref{sarh:transition-rate}) has the implication that when a polarisable particle moves though a medium with the field in its ground state, it can become excited.  In some sense, what we think of as the vacuum state is not stable to relative motion: both the field and the object can finish in an excited state.  The vacuum state of the field in the vicinity of a medium therefore abhors relative motion.

\begin{samepage}
\noindent\hrulefill
\paragraph*{Exercise:}  Through examining the frequency of a field \(\omega\) in two relatively moving frames derive the condition for \(\omega\) to take different signs in these two frames.  Can this occur in free space?  Given the discussion given in this section, how might we interpret this sign change?

\noindent\hrulefill\\
\end{samepage}

\subsubsection{Emission from the moving particle}

During the process of exciting the dipole the field gains momentum, which can be inferred from the Poynting vector (\ref{sarh:expect}).  The non--zero part of the integrated Poynting vector computed from (\ref{sarh:state}) is
\begin{multline}
	\int\langle\psi|\hat{S}_{x}|\psi\rangle dx=-\frac{1}{2\mu_{0}}\int\frac{dk'}{2\pi}\int_{0}^{\infty}d\omega'\int\frac{dk}{2\pi}\int_{0}^{\infty}d\omega c(k,\omega,t)c^{\star}(k',\omega',t)\\
	\times\int\langle0|\hat{C}_{\omega'}(k')\left[\frac{\partial \hat{A}_{z}}{\partial x}\frac{\partial \hat{A}_{z}}{\partial t}+\frac{\partial \hat{A}_{z}}{\partial t}\frac{\partial \hat{A}_{z}}{\partial x}\right]\hat{C}_{\omega}^{\dagger}(k)|0\rangle dx
\end{multline}
To make progress with this beastly object, we first work out the matrix element within the integrand, which is given by
\begin{multline}
	\int\langle0|\hat{C}_{\omega'}(k')\left[\frac{\partial \hat{A}_{z}}{\partial x}\frac{\partial \hat{A}_{z}}{\partial t}+\frac{\partial \hat{A}_{z}}{\partial t}\frac{\partial \hat{A}_{z}}{\partial x}\right]\hat{C}_{\omega}^{\dagger}(k)|0\rangle dx\\
	=-2\pi\delta(k-k')\mu_{0}^{2}\hbar k\mathcal{F}(\omega,\omega') g(k,\omega)g^{\star}(k,\omega')e^{-i(\omega-\omega')t}\label{sarh:me},
\end{multline}
where we have defined the quantity
\[
	\mathcal{F}(\omega,\omega')=\alpha(\omega)\alpha(\omega')\sqrt{\omega\omega'}(\omega+\omega').
\]
Notice that the matrix element (\ref{sarh:me}) is an odd function of \(k\), and that the function \(c(k,\omega,t)\) is independent of \(k\) when \(V=0\).  Therefore when \(V=0\) the total Poynting vector---which involves an integral over all values of $k$---is zero, and the coupling of the field to the particle does not result in any net power flow in the field.  For times \(t\to\infty\), the integrated Poynting vector takes a simple form, which is similar to the transition rate of the particle (\ref{sarh:transition-rate}):
\begin{equation}
	\int\langle\psi|\hat{S}_{x}|\psi\rangle dx\sim2\mu_{0}\beta^{2}\omega_{0}^{2}\int_{\omega_{0}/V}^{\infty}\frac{dk}{2\pi} k(Vk-\omega_{0})\text{Im}[g(k,Vk-\omega_{0})]^{2}\label{sarh:poynting-fric}
\end{equation}
To obtain this result, the identity \(\lim_{\lambda\to\infty}[\exp(i\lambda x)-1]/x=i\pi\delta(x)-\text{P}(1/x)\) was applied, and the terms involving the principal value, which are integrals over oscillatory functions, were dropped as being relatively small.  The quantity (\ref{sarh:poynting-fric}) is positive, meaning that radiation is emitted from the moving particle in the same direction as \(V\).  There is thus a force opposing the velocity of the particle, ultimately bringing it to rest.  This is the force of quantum friction, and vanishes smoothly as \(V\to0\), when the lower limit of the integration in (\ref{sarh:poynting-fric}) tends to infinity.
%
%
\subsection{Moving bodies in 3D macroscopic QED}
\par
The results in three dimensions are natural generalisations of the one dimensional formulae, and so the extension---although cumbersome---is really not fundamentally different from the above discussion.  For the sake of brevity we therefore give the theory without giving an enormous amount of explanation.  

In covariant form the earlier Lagrangian density (\ref{sarh:lf} --- \ref{sarh:li}) is
\begin{align}
	\mathscr{L}_{F}&=-\frac{1}{4\mu_{0}}F_{\mu\nu}F^{\mu\nu}\label{sarh:lf3d}\\
	\mathscr{L}_{I}&=\frac{c}{2}F_{\mu\nu}P^{\mu\nu}\label{sarh:li3d}\\
	\mathscr{L}_{R}&=\frac{1}{2}\int_{0}^{\infty}d\omega\left[\left(V^{\mu}\frac{\partial\vec{X}_{\omega}}{\partial x^{\mu}}\right)^{2}-\omega^{2}\vec{X}_{\omega}^{2}\right]\label{sarh:lr3d}
\end{align}
where the \emph{electromagnetic field tensor}\index{electromagnetic field tensor} is defined as~\cite{sarh:volume2}
\[
	F_{\mu\nu}=\partial_{\mu}A_{\nu}-\partial_{\nu}A_{\mu}\equiv\left(\begin{matrix}0&E_{x}/c&E_{y}/c&E_{z}/c\\-E_{x}/c&0&-B_{z}&B_{y}\\-E_{y}/c&B_{z}&0&-B_{x}\\-E_{z}/c&-B_{y}&B_{x}&0\end{matrix}\right)
\]
and the \emph{polarisation tensor}\index{polarisation tensor} as\cite{sarh:volume8}
\[
	P^{\mu\nu}=\gamma\left(\begin{matrix}0&P_{x}/\gamma&P_{y}&P_{z}\\-P_{x}/\gamma&0&VP_{y}&VP_{z}\\-P_{y}&-VP_{y}&0&0\\-P_{z}&-VP_{z}&0&0\end{matrix}\right).
\]
In this case the velocity of the body is taken along the \(x\)--axis and rest frame polarisation is defined as
\[
	\vec{P}=\int_{0}^{\infty}d\omega\,\alpha(\omega,\vec{x}-\vec{R})\vec{X}_{\omega}.
\]
Note that despite its covariant form, the Lagrangian is a function of the polarisation in the rest frame, and the quantity \(\vec{X}_{\omega}\) remains a three dimensional vector. To this Lagrangian, \(L_{0}=\int dx[\mathscr{L}_{F}+\mathscr{L}_{I}+\mathscr{L}_{R}]\), we add the kinetic energy of the centre of mass,
\begin{equation}
	L=\frac{1}{2}M\vec{V}^{2}+L_{0}\label{sarh:3dlag},
\end{equation}
and in the regime \(\gamma\sim1\) this expression forms the basis for the theory of radiation pressure.

\subsubsection{Computing the force}

From the Lagrangian (\ref{sarh:3dlag}) we can immediately calculate the electromagnetic force on a macroscopic body, varying the action with respect to \(\vec{R}\) and \(\vec{V}\)
\[
	\frac{d}{dt}\left(\frac{\partial L}{\partial\vec{V}}\right)=\frac{\partial L}{\partial\vec{R}},
\]
which gives
\begin{multline}
	\frac{d}{dt}\left[M\vec{V}+\int d^{3}\vec{x}\int_{0}^{\infty}d\omega\left(\vec{\nabla}\vec{\otimes}\vec{X}_{\omega}\right)\vec{\cdot}\left(\frac{\partial\vec{X}_{\omega}}{\partial t}+\vec{V}\vec{\cdot}\vec{\nabla}\vec{X}_{\omega}\right)-\int d^{3}\vec{x}\vec{P}\vec{\times}\vec{B}\right]\\
	=\frac{\partial}{\partial\vec{R}}\int d^{3}\vec{x}\vec{P}\vec{\cdot}\left(\vec{E}+\vec{V}\vec{\times}\vec{B}\right)\label{sarh:3dforce1}.
\end{multline}
This can be simplified through applying the equation of motion for the reservoir, \((\partial_{t}+\vec{V}\vec{\cdot}\vec{\nabla})^{2}\vec{X}_{\omega}+\omega^{2}\vec{X}_{\omega}=\alpha(\omega)[\vec{E}+\vec{V}\vec{\times}\vec{B}]\) (c.f. (\ref{sarh:mveqm2})), and after the application of a few vector identities we find the expression for the force (\ref{sarh:3dforce1}) becomes
\begin{equation}
	M\dot{\vec{V}}=\int d^{3}\vec{x}\left[\rho\vec{E}+\vec{j}\vec{\times}\vec{B}\right]\label{sarh:3dlf},
\end{equation}
which is the total Lorentz force on the body, with charge density
\[
	\rho=-\vec{\nabla}\vec{\cdot}\vec{P}
\]
and current density
\[
	\vec{j}=-\vec{V}(\vec{\nabla}\vec{\cdot}\vec{P})+\int_{0}^{\infty}\alpha(\omega,\vec{x}-\vec{R})\left(\frac{\partial}{\partial t}+\vec{V}\vec{\cdot}\vec{\nabla}\right)\vec{X}_{\omega}.
\]
To obtain this expression the reader might find the following formula useful: 
\[				(\vec{\nabla}\vec{\otimes}\vec{P})\vec{\cdot}\vec{V}\vec{\times}\vec{B}=[(
\vec{V}\vec{\times}\vec{B})\vec{\times}\vec{\nabla}]\vec{\times}\vec{P}
+(\vec{\nabla}\vec{\cdot}\vec{P})\vec{V}\vec{\times}\vec{B}.
\]
Applying the Maxwell equations \(\vec{\nabla}\vec{\cdot}\vec{E}=\rho/\varepsilon_{0}\) and \(\vec{\nabla}\vec{\times}\vec{B}=\mu_{0}\vec{j}+c^{-2}\dot{\vec{E}}\) the force (\ref{sarh:3dlf}) can be re--written as
\begin{equation}
	M\dot{\vec{V}}=\int d^{3}\vec{x}\left[\vec{\nabla}\vec{\cdot}\tens{T}-\frac{1}{c^{2}}\frac{\partial\vec{S}}{\partial t}\right]\label{sarh:force2}	
\end{equation}
where we have recovered the \emph{stress tensor}\index{stress tensor},
\[
	\tens{T}=\varepsilon_{0}\left[\vec{E}\vec{\otimes}\vec{E}+c^{2}\vec{B}\vec{\otimes}\vec{B}-\frac{1}{2}\tensunit\left(\vec{E}^{2}+c^{2}\vec{B}^{2}\right)\right],
\]
and the \emph{Poynting vector}\index{Poynting vector},
\[
	\vec{S}=\frac{1}{\mu_{0}}\vec{E}\vec{\times}\vec{B}.
\]
As we have already established in the one--dimensional case, the time average of the equation of motion (\ref{sarh:force2}) reproduces the known result from the classical theory of radiation pressure: the average force is given by the integral of the stress tensor over the surface of the body.  This comes as a natural consequence extending the Lagrangian of macroscopic QED to moving media.

\subsubsection{Quantum theory}

From (\ref{sarh:lf3d}--\ref{sarh:lr3d}) the canonical momenta of this system are
\begin{align}
	\vec{\Pi}_{\vec{A}}&=\varepsilon_{0}\left(\dot{\vec{A}}+\vec{\nabla}\varphi\right)-\int_{0}^{\infty}d\omega\,\alpha(\omega,\vec{x}-\vec{R})\vec{X}_{\omega}\nonumber\\
	\vec{\Pi}_{\vec{X}_{\omega}}&=\frac{\partial\vec{X}_{\omega}}{\partial t}+(\vec{V}\vec{\cdot}\vec{\nabla})\vec{X}_{\omega}
\end{align}
and
\[
	\boldsymbol{p}=\frac{\partial L}{\partial\vec{V}}=M\vec{V}+\vec{\mathcal{A}},
\]
where
\begin{equation}
	\vec{\mathcal{A}}=\int d^{3}\vec{x}\int_{0}^{\infty}d\omega\left[(\vec{\nabla}\vec{\otimes}\vec{X}_{\omega})\vec{\cdot}\vec{\Pi}_{\vec{X}_{\omega}}-\alpha(\omega,\vec{x}-\vec{R})\vec{X}_{\omega}\vec{\times}\vec{B}\right]\label{sarh:pseudoA},
\end{equation}
a quantity which again plays the role of an effective vector potential for the motion of the centre of mass.  In terms of these canonical variables the Hamiltonian is
\begin{align}
	H&=\vec{p}\vec{\cdot}\vec{V}+\int d^{3}\vec{x}\left[\dot{\vec{A}}\vec{\cdot}\vec{\Pi}_{\vec{A}}+\int_{0}^{\infty}d\omega\dot{\vec{X}}_{\omega}\vec{\cdot}\vec{\Pi}_{\vec{X}_{\omega}}\right]-L\nonumber\\
	&=\frac{(\vec{p}-\vec{\mathcal{A}})^{2}}{2M}+H_{0},
\end{align}
where the Lagrangian is given by (\ref{sarh:3dlag}), and \(H_{0}\) equals the stationary result (\ref{sarh:3dmacham}).  In quantum mechanics the Hamiltonian takes the same form, which --- as expected --- is very similar to the one dimensional result (\ref{sarh:qHV}),
\begin{equation}
	\hat{H}=\frac{\left(\hat{\vec{p}}-\vec{\hat{\mathcal{A}}}\right)^{2}}{2M}+\hat{H}_{0}\label{sarh:3dVham},
\end{equation}
where the operator \(\hat{\vec{\mathcal{A}}}\) only formally differs from (\ref{sarh:pseudoA}) because we must symmetrise the ordering of the operators
\begin{multline}
	\vec{\hat{\mathcal{A}}}=\int d^{3}\vec{x}\int_{0}^{\infty}d\omega\bigg\{\frac{1}{2}\left[(\vec{\nabla}\vec{\otimes}\hat{\vec{X}}_{\omega})\vec{\cdot}\hat{\vec{\Pi}}_{X_{\omega}}+\hat{\vec{\Pi}}_{X_{\omega}}\vec{\cdot}(\hat{\vec{X}}_{\omega}\vec{\otimes}\overleftarrow{\vec{\nabla}})\right]\\
	-\alpha(\omega,\vec{x}-\hat{\vec{R}})\hat{\vec{X}}_{\omega}\vec{\times}\hat{\vec{B}}\bigg\}\label{sarh:3dGop}.
\end{multline}
The Hamiltonian (\ref{sarh:3dVham}) describes the quantum mechanical motion of a polarisable body in the electromagnetic field. The quantity \(\hat{H}_{0}\) is equal to the Hamiltonian for the field and medium for a body at rest (\ref{sarh:3dmacham}), at a position determined by \(\hat{\vec{R}}\).  The results for a stationary body can be reclaimed if we take the limit \(M\to\infty\).

\subsubsection{Quantum forces}

Our main concern is computing quantum vacuum forces, so the first thing is to determine the average acceleration of the centre of mass in response to a quantum state of the electromagnetic field.  Both the time derivative of the centre of mass operator, \(\hat{\vec{R}}\), and its acceleration are given by expressions that are formally identical to the classical ones, with the velocity given by
\begin{equation}
	\frac{d\hat{\vec{R}}}{dt}=\frac{i}{\hbar}\left[\hat{H},\hat{\vec{R}}\right]=\frac{(\hat{\vec{p}}-\hat{\vec{\mathcal{A}}})}{M}\equiv\hat{\vec{V}}\label{sarh:3dV},
\end{equation}
where we have applied \([\hat{\vec{R}},\hat{\vec{p}}]=i\hbar\tensunit\).  The acceleration is then equal to
\begin{align}
	M\frac{d^{2}\hat{\vec{R}}}{dt^{2}}&=\frac{i}{\hbar}\left[\hat{H},\hat{\vec{p}}-\hat{\vec{\mathcal{A}}}\right]\nonumber\\
	&=\int d^{3}\vec{x}\left[\hat{\rho}\hat{\vec{E}}+\hat{\vec{j}}\vec{\times}\hat{\vec{B}}\right]\label{sarh:qforcea},
\end{align}
where the charge and current density operators are
\begin{align*}
	\hat{\rho}&=-\vec{\nabla}\vec{\cdot}\hat{\vec{P}}\\
	\hat{\vec{j}}&=\int_{0}^{\infty}\alpha(\omega,\vec{x}-\vec{R})\hat{\vec{\Pi}}_{X_{\omega}}+\frac{1}{2}\left(\hat{\vec{V}}\hat{\rho}+\hat{\rho}\hat{\vec{V}}\right).
\end{align*}

\begin{samepage}
\noindent\hrulefill
\paragraph*{Exercise:}Fill in the steps between the two lines of (\ref{sarh:qforcea}).\\
\emph{Hint}: Don't forget that in this vector case \(\hat{\vec{p}}-\hat{\vec{\mathcal{A}}}\) does not commute with \((\hat{\vec{p}}-\hat{\vec{\mathcal{A}}})^{2}\).

\noindent\hrulefill\\
\end{samepage}

The right hand side of the equation of motion (\ref{sarh:qforcea}) is formally the same as the classical formula (\ref{sarh:3dlf}), and as the field operators obey the classical Maxwell equations, we can also re--express this force in terms of the fields alone, finding the operator analogue of our classical expression (\ref{sarh:force2})
\[
	M\frac{d^{2}\hat{\vec{R}}}{dt^{2}}=\int d^{3}\vec{x}\left[\vec{\nabla}\vec{\cdot}\hat{\tens{T}}-\frac{1}{c^{2}}\frac{\partial\hat{\vec{S}}}{\partial t}\right],
\]
where we have defined the \emph{stress tensor} operator
\begin{equation}
	\hat{\tens{T}}=\varepsilon_{0}\left[\hat{\vec{E}}\vec{\otimes}\hat{\vec{E}}+c^{2}\hat{\vec{B}}\vec{\otimes}\hat{\vec{B}}-\frac{1}{2}\tensunit\left(\hat{\vec{E}}^{2}+c^{2}\hat{\vec{B}}^{2}\right)\right]\label{sarh:st3d}
\end{equation}
and the Poynting vector operator
\[
	\hat{\vec{S}}=\frac{1}{2\mu_{0}}\left[\hat{\vec{E}}\vec{\times}\hat{\vec{B}}-\hat{\vec{B}}\vec{\times}\hat{\vec{E}}\right].
\]
We have thus found, from first principles, that the centre of mass of a body obeys the operator equivalent of the classical theory of radiation pressure.  For a stationary localised object at position \(\vec{R}_{0}\) with the field and medium in the ground state --- as we considered in section \ref{sarh:vacuum-force} --- the average value of the force on a body is given by
\begin{equation}
	M\frac{d^{2}\langle\hat{\vec{R}}\rangle}{dt^{2}}=\int_{\partial V} d \vec{s}\vec{\cdot}\langle0_{\vec{R}_{0}}|\hat{\tens{T}}|0_{\vec{R}_{0}}\rangle\label{sarh:qforce},
\end{equation}
where \(\partial V\) stands for the surface of the body, and
\begin{equation}
	\langle0_{\vec{R}_{0}}|\hat{\tens{T}}|0_{\vec{R}_{0}}\rangle=\frac{\hbar}{\pi}\lim_{\vec{x}_{1}\to\vec{x}_{2}}\int_{0}^{\infty}d\omega\,\bigg\{\tens{\theta}(\vec{x}_{1},\vec{x}_{2},\omega)-\frac{1}{2}\tensunit\text{Tr}\left[\tens{\theta}(\vec{x}_{1},\vec{x}_{2},\omega)\right]\bigg\}
\end{equation}
with the rank two tensor $\tens{\theta}$ defined as
\begin{equation}
	\tens{\theta}(\vec{x}_{1},\vec{x}_{2},\omega)=\frac{\omega^{2}}{c^{2}}\text{Im}\left[\tens{G}^{(S)}(\vec{x}_{1},\vec{x}_{2},\omega)\right]+\vec{\nabla}_{1}\vec{\times}\text{Im}\left[\tens{G}^{(S)}(\vec{x}_{1},\vec{x}_{2},\omega)\right]\vec{\times}\overleftarrow{\vec{\nabla}}_{2}\label{sarh:Ttensor}.
\end{equation}
The superscript `\((S)\)' implies the scattered part of the Green function, as discussed in section~\ref{sarh:vacuum-force}.  To obtain this expression for the force we used the field operators given by (\ref{sarh:3dfop}), as well as the integral identity for Green functions given by (\ref{sarh:green-identity2}).  Equation (\ref{sarh:qforce}) establishes that the ground state electromagnetic field imparts an acceleration to a body that, on average, is equal to the expression used in Lifshitz theory~\cite{sarh:volume9}, in the case of a body surrounded by vacuum.  We have recovered Lifshitz theory as a natural consequence of self--consistently applying macro--QED to moving media.  But this only turns out to be a special case, and there is no restriction on what the state of the system is, besides the assumption that macroscopic electromagnetism is valid.
%
%
\subsection{Quantum friction between sliding plates\index{quantum friction}}

\par
To conclude the tutorial, we'll apply the above theory to the phenomenon of quantum friction, which was already partially discussed in section~\ref{sarh:qfsec}.  In our 1D treatment the friction phenomenon was inferred from the behaviour of a polarisable particle moving at a constant velocity through a medium.  We now aim to show directly that when two separated bodies slide past one another, there is a force that serves to bring them to relative rest even for perfectly smooth bodies at zero temperature.  As we shall see, this is a situation when the standard theory of Casimir forces cannot be applied.  Indeed, there has been some controversy over the existence of this force that came from comparing results derived within the formalism of Lifshitz theory to results derived within a different framework~\cite{sarh:pendry1997,sarh:leonhardt2009}.  Using macro--QED we shall show that the usual fluctuation--dissipation theorem that underlies Lifshitz theory is not applicable to bodies in relative motion, and we find an extra term in the correlation function which turns out to be the source of the controversy.  Using macro--QED we shall show that it is then possible to reclaim the expression for the frictional force given by J. B. Pendry~\cite{sarh:pendry1997}.
%
%
\begin{figure}[h!]
	\begin{center}
	\includegraphics[width=5.3cm]{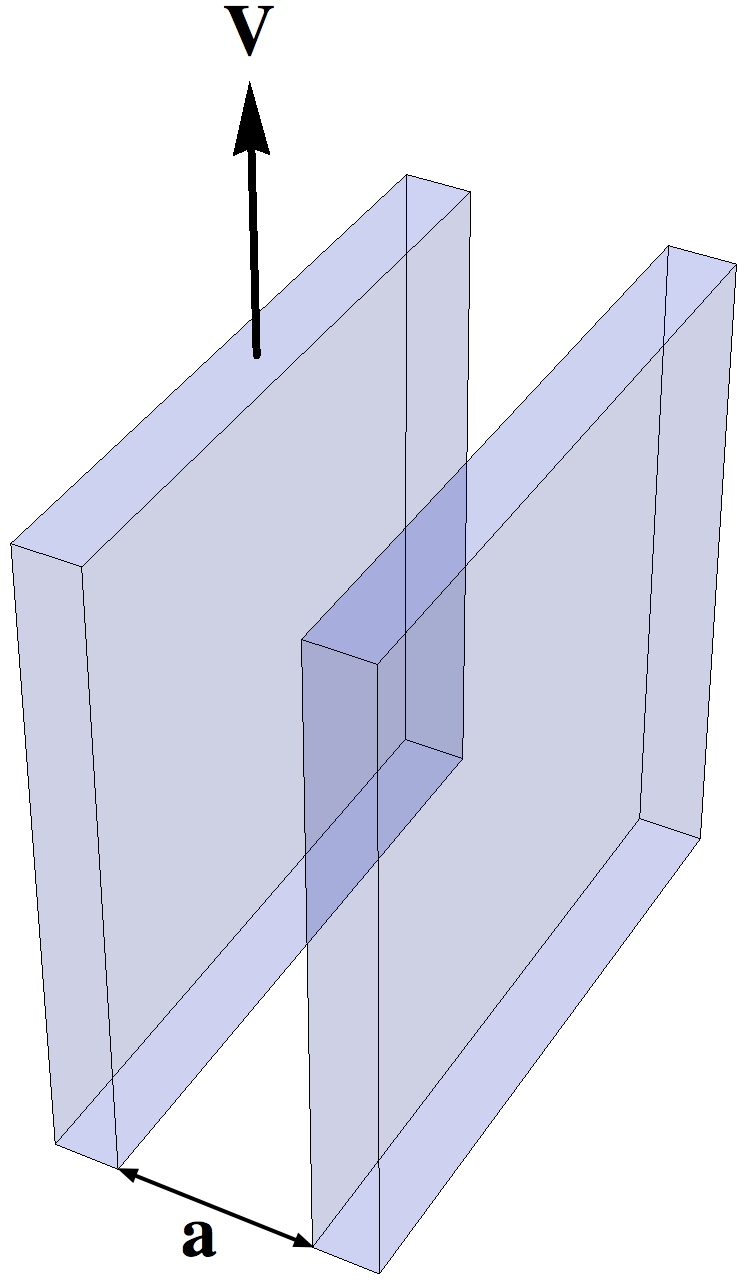}
	\caption{Two planar materials are separated by a distance \(a\) and slide relative to one another with velocity \(\vec{V}\).  The ground state field in the gap between the two bodies can be thought of as being generated by currents within each of them.  The interaction between these currents serves to bring the relative motion to zero, a phenomenon known as quantum friction.\label{sarh:qmfriction-fig}}
	\end{center}
\end{figure}

Consider two semi--infinite planar media separated by a distance \(a\), with one of the bodies (say the one on the left) moving at a velocity \(\vec{V}\) lying in the \(y\)--\(z\) plane (see figure~\ref{sarh:qmfriction-fig}).  We assume that the bodies are massive enough that the velocity operator can be replaced with the vector \(\vec{V}\).  We have already established in equation (\ref{sarh:jop}) that the current operator associated with excitations of electrical current in the stationary body is given by
\begin{equation}
	\hat{\vec{j}}_{R}(\vec{x},t)=\frac{\partial\hat{\vec{P}}_{R}(\vec{x},t)}{\partial t}\label{sarh:jr},
\end{equation}
where
\begin{equation}
	\hat{\vec{P}}_{L,R}(\vec{x},t)=\int_{0}^{\infty}d\omega\int\frac{d^{2}\vec{k}_{\parallel}}{(2\pi)^{2}}\sqrt{\frac{\hbar}{2\omega}}\alpha_{L,R}(\omega,x)\hat{\vec{C}}_{\omega}(x,\vec{k}_{\parallel})e^{i(\vec{k}_{\parallel}\vec{\cdot}\vec{x}-\omega t)}+\text{h.c.}\label{sarh:plr}
\end{equation}
and
\[
	\alpha_{L,R}(\omega,x)=\alpha(\omega)\begin{cases}\Theta(-x)&L\\\Theta(x-a)&R.\end{cases}
\]
Meanwhile the operator for current in the moving plate takes the form of a Lorentz transformation of the rest frame value
\begin{align}
	\hat{\vec{j}}_{L}(\vec{x},t)&=\frac{\partial\hat{\vec{P}}_{L}(\vec{x}-\vec{V}t,t)}{\partial t}-\vec{\nabla}\vec{\times}\left(\vec{V}\vec{\times}\hat{\vec{P}}_{L}(\vec{x}-\vec{V}t,t)\right)\nonumber\\
	&\sim\frac{\partial\hat{\vec{P}}_{L}(\vec{x}-\vec{V}t,t)}{\partial t}\label{sarh:jl},
\end{align}
an expression which is valid only to leading order and neglects all relativistic effects, barring the Doppler shift\index{Doppler effect}.  In the final line we have neglected the curl of a quantity, which is an entirely transverse contribution to the current density.  This is valid because in the end we shall take a low velocity limit where the plates are closely spaced.  In such a limit only the longitudinal (non-retarded) components of the field are relevant.

\begin{samepage}
\noindent\hrulefill
\paragraph{Exercise:}Show that --- for low velocities --- the current transforms as (\ref{sarh:jl}).

\noindent\hrulefill\\
\end{samepage}

The total current in the system is given by the sum of these two contributions, \(\hat{\vec{j}}=\hat{\vec{j}}_{L}+\hat{\vec{j}}_{R}\), and this is the source of the electric field
\begin{equation}
	\hat{\vec{E}}(\vec{x},t)=-\mu_{0}\int d^{3}\vec{x}'\int_{-\infty}^{t}dt'\tens{G}(\vec{x},\vec{x}',t-t')\vec{\cdot}\frac{\partial\hat{\vec{j}}(\vec{x}',t')}{\partial t'},\label{sarh:electric-field-mm}
\end{equation}
where the Green function in the above formula is for the whole system and is written in the time rather than frequency domain.  From this expression for the electric field operator (\ref{sarh:electric-field-mm}), we can calculate the stress tensor~\footnote{The magnetic field operator can be inferred from the Maxwell equation \(\vec{\nabla}\vec{\times}\hat{\vec{E}}=-\frac{\partial\hat{\vec{B}}}{\partial t}\).} (\ref{sarh:st3d}).  The expression for the force (\ref{sarh:qforce}) shows that only the off diagonal components of the stress tensor are relevant for the lateral (frictional) force, which is of interest here.  To determine these off diagonal components we evaluate the electric field correlation function,
\begin{multline}
	\langle0|\hat{\vec{E}}(\vec{x},t)\vec{\otimes}\hat{\vec{E}}(\vec{x}',t)|0\rangle=\mu_{0}^{2}\int d^{3}\vec{x}_{1}\int_{-\infty}^{t}dt_{1}\int d^{3}\vec{x}_{2}\int_{-\infty}^{t}dt_{2}\\
	\times\tens{G}(\vec{x},\vec{x}_{1},t-t_{1})\vec{\cdot}\langle0|\frac{\partial\hat{\vec{j}}(\vec{x}_{1},t_{1})}{\partial t_{1}}\vec{\otimes}\frac{\partial\hat{\vec{j}}(\vec{x}_{2},t_{2})}{\partial t_{2}}|0\rangle\vec{\cdot}\tens{G}^{T}(\vec{x}',\vec{x}_{2},t-t_{2})\label{sarh:ecqf}.
\end{multline}
The correlation in the electric field is thus determined by the ground state correlation in the electrical current within the two plates, which is modified by the relative motion.  Using their representation in terms of the creation and annihilation operators (\ref{sarh:jr}--\ref{sarh:jl}) we find the vacuum correlation between the electrical currents is given by
\begin{multline}
	\langle0|\frac{\partial\hat{\vec{j}}(\vec{x}_{1},t_{1})}{\partial t_{1}}\tprod\frac{\partial\hat{\vec{j}}(\vec{x}_{2},t_{2})}{\partial t_{2}}|0\rangle=\tensunit\delta(x_{1}-x_{2})\int_{0}^{\infty}d\omega\int\frac{d^{2}\vec{k}_{\parallel}}{(2\pi)^{2}}\frac{\hbar\varepsilon_{0}}{\pi}\text{Im}[\varepsilon(\omega)]e^{i\vec{k}_{\parallel}\vec{\cdot}(\vec{x}_{1}-\vec{x}_{2})}\\
	\times\bigg\{\Theta(x_{1}-a)\omega^{4}e^{-i\omega(t_{1}-t_{2})}+\Theta(-x_{1})\omega_{+}^{4}e^{-i\omega_{+}(t_{1}-t_{2})}\bigg\}\label{sarh:frictioncorr}
\end{multline}
where \(\omega_{\pm}=\omega\pm\vec{V}\vec{\cdot}\vec{k}_{\parallel}\), and \(\vec{k}_{\parallel}\) is a wave--vector lying in the \(y\)--\(z\) plane.  Inserting the electric current correlation function (\ref{sarh:frictioncorr}) into (\ref{sarh:ecqf}) gives us the electric field correlation function written without reference to the operators
\begin{multline}
	\langle0|\hat{\vec{E}}(\vec{x},t)\tprod\hat{\vec{E}}(\vec{x}',t)|0\rangle=\frac{\hbar\mu_{0}}{\pi c^{2}}\int_{0}^{\infty}d\omega\int\frac{d^{2}\vec{k}_{\parallel}}{(2\pi)^{2}}\text{Im}[\varepsilon(\omega)]e^{i\vec{k}_{\parallel}\vec{\cdot}(\vec{x}-\vec{x}')}\\
	\times\bigg\{\omega^{4}\int_{a}^{\infty}dx_{1}\tens{G}(x,x_{1},\vec{k}_{\parallel},\omega)\vec{\cdot}\tens{G}^{\dagger}(x',x_{1},\vec{k}_{\parallel},\omega)\\
	+\omega_{+}^{4}\int_{-\infty}^{0}dx_{1}\tens{G}(x,x_{1},\vec{k}_{\parallel},\omega_{+})\vec{\cdot}\tens{G}^{\dagger}(x',x_{1},\vec{k}_{\parallel},\omega_{+})\bigg\}\label{sarh:nearlyecorr}
\end{multline}
where \(\tens{G}^{\dagger}=(\tens{G}^{T})^{\star}\).\\[10pt]

{\footnotesize
\noindent
\textbf{Integral identities for the Green function}:  To this order we are neglecting all effects of the motion except the Doppler shift within the dispersion of the medium.  Therefore the differential equation satisfied by the Green function is
\begin{equation}
	\vec{\nabla}\vec{\times}\vec{\nabla}\vec{\times}\tens{G}(x,x',\vec{k}_{\parallel},\omega)-\frac{\omega^{2}}{c^{2}}[\varepsilon(\omega_{-})\Theta(-x)+\varepsilon(\omega)\Theta(x-a)]\tens{G}(x,x',\vec{k}_{\parallel},\omega)=\tensunit\delta(x-x')\label{sarh:green-eqn-qf1},
\end{equation}
where \(\vec{\nabla}=\vec{e}_{x}\partial_{x}+i\vec{k}_{\parallel}\), and \(\omega_{\pm}=\omega\pm\vec{V}\vec{\cdot}\vec{k}_{\parallel}\).  Meanwhile the Hermitian conjugate of the Green function obeys
\begin{equation}
	\vec{\nabla}\vec{\times}\vec{\nabla}\vec{\times}\tens{G}^{\dagger}(x',x,\vec{k}_{\parallel},\omega)-\frac{\omega^{2}}{c^{2}}[\varepsilon^{\star}(\omega_{-})\Theta(-x)+\varepsilon^{\star}(\omega)\Theta(x-a)]\tens{G}^{\dagger}(x',x,\vec{k}_{\parallel},\omega)=\tensunit\delta(x-x')\label{sarh:green-eqn-qf2}
\end{equation}
Multiplying (\ref{sarh:green-eqn-qf2}) on the left by the Green function, \(\tens{G}(x,x'',\vec{k}_{\parallel},\omega)\), integrating over \(x\), and then subtracting the Hermitian conjugate of the resulting expression with \(x'\) and \(x''\) reversed, we obtain a generalisation of (\ref{sarh:green-identity2}),
\begin{multline}
	\frac{\tens{G}(x',x'',\vec{k}_{\parallel},\omega)-\tens{G}^{\dagger}(x'',x',\vec{k}_{\parallel},\omega)}{2i}=\frac{\omega^{2}}{c^{2}}\bigg[\text{Im}[\varepsilon(\omega_{-})]\int_{-\infty}^{0} dx\tens{G}(x',x,\vec{k}_{\parallel},\omega)\vec{\cdot}\tens{G}^{\dagger}(x'',x,\vec{k}_{\parallel},\omega)\\
	+\text{Im}[\varepsilon(\omega)]\int_{a}^{\infty} dx\tens{G}(x',x,\vec{k}_{\parallel},\omega)\vec{\cdot}\tens{G}^{\dagger}(x'',x,\vec{k}_{\parallel},\omega)\bigg].\label{sarh:green-identity-qf}
\end{multline}
For integration over only the region \(x<0\) we have instead
\begin{multline}
	\frac{\omega^{2}}{c^{2}}\text{Im}[\varepsilon(\omega_{-})]\int_{-\infty}^{0}dx\tens{G}(x',x,\vec{k}_{\parallel},\omega)\vec{\cdot}\tens{G}^{\dagger}(x'',x,\vec{k}_{\parallel},\omega)=\frac{\tens{G}(x',x'',\vec{k}_{\parallel},\omega)-\tens{G}^{\dagger}(x'',x',\vec{k}_{\parallel},\omega)}{2i}\\
	-\frac{1}{2i}\left[\tens{G}(x',0,\vec{k}_{\parallel},\omega)\vec{\cdot}\vec{e}_{x}\vec{\times}\vec{\nabla}\vec{\times}\tens{G}^{\dagger}(x'',0,\vec{k}_{\parallel},\omega)-\text{h.c.}\right]\label{sarh:surface-term}.
\end{multline}
}

Applying (\ref{sarh:green-identity-qf}) to (\ref{sarh:nearlyecorr}), gives us the electric field correlation function, and after taking the limit, \(\vec{x}\to\vec{x}'\), we obtain the electric contribution to the stress tensor
\begin{multline}
	\langle0|\hat{\vec{E}}(\vec{x},t)\vec{\otimes}\hat{\vec{E}}(\vec{x},t)|0\rangle=\frac{\hbar\mu_{0}}{\pi}\int\frac{d^{2}\vec{k}_{\parallel}}{(2\pi)^{2}}\int_{0}^{\infty}d\omega\,\omega^{2}\text{Im}[\bar{\tens{G}}^{(S)}(x,x,\vec{k}_{\parallel},\omega)]\\
	-\int_{-\infty}^{0}dx_{1}\int\frac{d^{2}\vec{k}_{\parallel}}{(2\pi)^{2}}\int_{0}^{\vec{V}\vec{\cdot}\vec{k}_{\parallel}}d\omega \bar{\tens{Q}}(x,x_{1},\vec{k}_{\parallel},\omega)\label{sarh:lifqf},
\end{multline}
where we have introduced symmetrised quantities, e.g. \(\bar{\tens{G}}=\frac{1}{2}[\tens{G}+\tens{G}^{T}]\), which makes explicit the independence of the right hand side from the order of the electric field operators.  The quantity \(\tens{Q}\) within the correlation function (\ref{sarh:lifqf}) is given by
\begin{equation}
	\tens{Q}(x,x_{1},\vec{k}_{\parallel},\omega)=\frac{\hbar\mu_{0}}{\pi}\frac{\omega^{4}}{c^{2}}\text{Im}[\varepsilon(\omega_{-})]\tens{G}(x,x_{1},\vec{k}_{\parallel},\omega)\vec{\cdot}\tens{G}^{\dagger}(x,x_{1},\vec{k}_{\parallel},\omega).
\end{equation}
The result (\ref{sarh:lifqf}) is that which enters Lifshitz theory (\ref{sarh:Ttensor}), plus an additional velocity dependent term equal to an integral over low frequency excitations within the moving plate.  The new term \(\tens{Q}\) comes from an imbalance between the frequency spectrum of the electrical current in the stationary plate and the one in motion, which in turn is due to the Doppler effect.  This is the origin of the friction between the plates.  The imbalance of frequencies is similar to the effect discussed in section~\ref{sarh:qfsec}, where the Doppler effect causes a sign change of some of the frequencies between reference frames, which can slow a moving particle.

Integrating the term on the second line of (\ref{sarh:lifqf}) over the moving plate and applying the result (\ref{sarh:surface-term}) results in two terms,
\begin{multline}
	\int_{-\infty}^{0}dx_{1} \tens{Q}(x,x_{1},\vec{k}_{\parallel},\omega)=\frac{\hbar\mu_{0}\omega^{2}}{\pi}\frac{\tens{G}(x,x,\vec{k}_{\parallel},\omega)-\tens{G}^{\dagger}(x,x,\vec{k}_{\parallel},\omega)}{2i}\\
	-\frac{\hbar\mu_{0}\omega^{2}}{2\pi i}\left\{\tens{G}(x,0,\vec{k}_{\parallel},\omega)\vec{\cdot}\vec{e}_{x}\vec{\times}\left(\vec{\nabla}\vec{\times}\tens{G}^{\dagger}(x,0,\vec{k}_{\parallel},\omega)\right)-\text{h.c.}\right\}\label{sarh:flux}
\end{multline}
where the curl in the second line is taken with respect to the second spatial coordinate.  The term on the second line came from an integration by parts, and can be thought of as a flux of momentum leaving the moving plate (it takes the form of an electric field crossed with a magnetic field, like the Poynting vector).  This is the term responsible for the frictional force between the plates, which we now examine in detail, using the explicit form for the Green function.\\[10pt]

{\footnotesize
\textbf{The Green function between relatively moving plates:}  In empty space the Green function satisfies 
\[
	\boldsymbol{\nabla}\times\boldsymbol{\nabla}\times\tens{G}_0-k_0^2\tens{G}_0=\tensunit\delta^{(3)}(\boldsymbol{x}-\boldsymbol{x}')
\]
When written in terms of \(x\), \(x'\), \(\vec{k}_{\parallel}\) and \(\omega\) this equals
\[
	\tens{G}_{0}(x,x',\vec{k}_{\parallel},\omega)=\left[\tensunit-\frac{\vec{k}_{\pm}\tprod\vec{k}_{\pm}}{k_{0}^{2}}\right]\frac{i e^{i \sqrt{k_{0}^{2}-k_{\parallel}^{2}}|x-x'|}}{2\sqrt{k_{0}^{2}-k_{\parallel}^{2}}}.
\]
where \(k_{0}=\omega/c\) and
\[
	\vec{k}_{\pm}=\text{sign}(x-x')\vec{e}_{x}\sqrt{k_{0}^{2}-k_{\parallel}^{2}}+\vec{k}_{\parallel},
\]
which is simply the 1D Green function (\ref{sarh:1dgreen}) with a bit of dressing to take account of polarisation and angle of incidence.  For realistic sliding velocities (on the order of metres per second), \(\tens{Q}\) is evaluated in a regime where \(k_{\parallel}\gg k_{0}\) and the field exponentially dies away from the source\footnote{This is known as the \emph{electrostatic}, or \emph{non--retarded} limit by those working on the physics of the electromagnetic field close to surfaces.  See, e.g.~\cite{sarh:mayergoyz2013}.}.  In this regime only the first few reflections (we consider two) of the field from the two plates contribute significantly to the Green function.  Moreover when the permeability \(\mu=1\) the Fresnel reflection coefficient\index{Fresnel reflection coefficient} for \(s\)--polarised radiation is zero in this limit\footnote{See e.g.~\cite{sarh:volume8}}.  Therefore the Green function can be approximated by
\begin{multline}
	\tens{G}(x,x',\vec{k}_{\parallel},\omega)\sim\tens{G}_{0}(x,x',\vec{k}_{\parallel},\omega)+\frac{1}{2\kappa}\bigg[\vec{u}\vec{\otimes}\vec{u}^{\star}\,r_{p}(\omega_{-})e^{-\kappa(x+x')}+\vec{u}^{\star}\vec{\otimes}\vec{u}\,r_{p}(\omega)e^{-\kappa(2a-x-x')}\\[5pt]
	+\vec{u}^{\star}\vec{\otimes}\vec{u}^{\star}\,r_{p}(\omega)r_{p}(\omega_{-})e^{-\kappa(2a+x'-x)}+\vec{u}\vec{\otimes}\vec{u}\,r_{p}(\omega)r_{p}(\omega_{-})e^{-\kappa(2a+x-x')}\bigg],\\\label{sarh:gfea}
\end{multline}
where \(\kappa=\sqrt{k_{\parallel}^{2}-k_{0}^{2}}\), \(r_{p}(\omega)\) is the \(p\)--polarised reflection coefficient for a body at rest, and the \(p\)--polarised and \(s\)--polarised unit vectors are respectively given by
\begin{align}
	\vec{u}&=\frac{1}{k_{0}k_{\parallel}}[k_{\parallel}^{2}\vec{e}_{x}- i\kappa\vec{k}_{\parallel}]\nonumber\\
	\vec{v}&=\frac{1}{k_{\parallel}}\vec{e}_{x}\vec{\times}\vec{k}_{\parallel}.
\end{align}
The curl of the Hermitian conjugate of the Green function (\ref{sarh:gfea}) with respect to the second index is
\begin{multline}
	\vec{\nabla}'\vec{\times}\tens{G}^{\dagger}(x,x',\vec{k}_{\parallel},\omega)\sim\vec{\nabla}'\vec{\times}\tens{G}_{0}(x,x',\vec{k}_{\parallel},\omega)-\frac{i k_{0}}{2\kappa}\vec{v}\tprod\bigg[\vec{u}^{\star}r_{p}^{\star}(\omega_{-})e^{-\kappa(x+x')}+\vec{u}r_{p}^{\star}(\omega)e^{-\kappa(2a-x-x')}\\[5pt]
	+\vec{u}r_{p}^{\star}(\omega)r_{p}^{\star}(\omega_{-})e^{-\kappa(2a+x'-x)}+\vec{u}^{\star}r_{p}^{\star}(\omega)r_{p}^{\star}(\omega_{-})e^{-\kappa(2a+x-x')}\bigg].\label{sarh:gfcurl}
\end{multline}
\\[10pt]
}
Inserting expressions (\ref{sarh:gfea}) and (\ref{sarh:gfcurl}) for the Green function into the surface term in (\ref{sarh:flux}) we find
\begin{multline}
	\frac{1}{2i}\vec{e}_{x}\vec{\cdot}\left[\tens{G}(x,0,\vec{k}_{\parallel},\omega)\vec{\cdot}\vec{e}_{x}\vec{\times}(\vec{\nabla}\vec{\times}\tens{G}^{\dagger}(x,0,\vec{k}_{\parallel},\omega))-\text{h.c.}\right]\vec{\cdot}\vec{e}_{y}\\
	=\frac{k_{y}e^{-2\kappa a}}{k_{0}^{2}}\text{Im}[r_{p}(\omega)]\text{Im}[r_{p}(\omega_{-})]\label{sarh:Gresult}
\end{multline}
where terms of order \(\exp(-4\kappa a)\) and smaller have been neglected and only the symmetric part of the tensor has been retained.  The \(x\)--\(y\) component of the electric field correlation function (\ref{sarh:lifqf}) is
\begin{multline}
	\langle0|\hat{E}_{x}(\vec{x},t)\hat{E}_{y}(\vec{x},t)|0\rangle=\frac{\hbar\mu_{0}}{\pi}\int\frac{d^{2}\vec{k}_{\parallel}}{(2\pi)^{2}}\int_{V k_{y}}^{\infty}d\omega\omega^{2}\text{Im}[\bar{G}^{(S)}_{xy}(x,x,\vec{k}_{\parallel},\omega)]+\\
	\frac{\hbar\mu_{0}}{2\pi i}\int\frac{d^{2}\vec{k}_{\parallel}}{(2\pi)^{2}}\int_{0}^{Vk{y}}d\omega\omega^{2}\vec{e}_{x}\vec{\cdot}\left[\tens{G}(x,0,\vec{k}_{\parallel},\omega)\vec{\cdot}\vec{e}_{x}\vec{\times}\left(\vec{\nabla}\vec{\times}\tens{G}^{\dagger}(x,0,\vec{k}_{\parallel},\omega)\right)-\text{h.c.}\right]\vec{\cdot}\vec{e}_{y}\label{sarh:ecf},
\end{multline}
where only the symmetric part of the tensor on the second line is included.  In this `electrostatic' regime the magnetic correlation function does not contribute to the stress\footnote{See~\cite{sarh:pendry1997}.}.  Furthermore it was shown in~\cite{sarh:leonhardt2009} that the first term on the right of (\ref{sarh:ecf}) is zero, although we shall not prove this here.  Therefore the frictional force is \(\varepsilon_{0}\) times the second term in (\ref{sarh:ecf}) which after we apply (\ref{sarh:Gresult}) is
\begin{equation}
	\langle0|\hat{T}_{xy}|0\rangle=\frac{\hbar}{\pi}\int\frac{d^{2}\vec{k}_{\parallel}}{(2\pi)^{2}}\int_{0}^{V k_{y}}d\omega k_{y}\text{Im}[r_{p}(\omega)]\text{Im}[r_{p}(\omega_{-})]e^{-2k_{\parallel}a}\label{sarh:friction-final},
\end{equation}
where we have taken the limit \(k_{\parallel}\gg k_{0}\).  The above off diagonal element of the stress tensor is the limiting expression found by Volokitin and Persson~\cite{sarh:volokitin2008} and Pendry~\cite{sarh:pendry2010}.  As we already mentioned, this limit is appropriate for low sliding velocities (relative to the speed of light), and we can see that the frictional force is determined by the imaginary parts of the reflection coefficients at low frequencies and large wave--vectors.  The integral over \(\omega\) is such that \(\omega_{-}\leq0\) and therefore \(\text{Im}[r_{p}(\omega_{-})]\leq0\) and the stress is negative.  This means that the force on the left hand body acts against its motion, and the force on the right hand body acts in the opposite direction; i.e. the relative motion of the bodies is being reduced.  Roughly speaking, for materials with a large degree of dissipation this frictional force is large.  The vacuum field close to a dissipative body can thus be thought of as being like a viscous fluid that inhibits motion parallel to the surface.

\begin{samepage}
\noindent\hrulefill
	\paragraph{Exercise:}Show that the reflection coefficient for \(p\)--polarised radiation
	\[
		r_{p}=\frac{\varepsilon(\omega)\sqrt{k_{0}^{2}-k_{\parallel}^{2}}-\sqrt{\varepsilon(\omega)k_{0}^{2}-k_{\parallel}^{2}}}{\varepsilon(\omega)\sqrt{k_{0}^{2}+k_{\parallel}^{2}}+\sqrt{\varepsilon(\omega)k_{0}^{2}-k_{\parallel}^{2}}}
	\]
	approaches \((\varepsilon(\omega)-1)/(\varepsilon(\omega)+1)\) as \(k_{\parallel}/k_{0}\to\infty\).  Using this result show that in the same limit
	\[
		\text{Im}[r_{p}(\omega)]\to\frac{2\text{Im}[\varepsilon(\omega)]}{|\varepsilon(\omega)+1|^{2}}.
	\]
	For the simple case of a constant conductivity \(\sigma\) the permittivity is \(\varepsilon(\omega)=\varepsilon_{b}+i\sigma/\omega\varepsilon_{0}\) (\(\varepsilon_{b}\) assumed constant).  For this case show that the stress tensor (\ref{sarh:friction-final}) is given by
	\[
		\langle0|\hat{T}_{xy}|0\rangle=\frac{4\sigma^{2}\hbar}{\pi\varepsilon_{0}^{2}}\int\frac{d^{2}\vec{k}_{\parallel}}{(2\pi)^{2}}\int_{0}^{V k_{y}}d\omega k_{y}\frac{\omega\omega_{-}}{|i\sigma/\varepsilon_{0}+(1+\varepsilon_{b})\omega|^{2}|i\sigma/\varepsilon_{0}+(1+\varepsilon_{b})\omega_{-}|^{2}}e^{-2k_{\parallel}a}.
	\]
	Using your favourite software or programming language numerically evaluate this integral as a function of \(V\) and \(a\) and plot the results.
	
\noindent\hrulefill\\
\end{samepage}

We complete this tutorial having developed the theory of quantum friction and quantum forces, all selfconsistently within the formalism of macroscopic QED.  This formalism is a very general way of treating the quantum mechanics of macroscopic bodies and the electromagnetic field and may be applied to any problem in Casimir physics.  Not only does this justify the formulae of Lifshitz theory on the basis of a complete quantum mechanical theory, but it opens up the possibility of exploring new effects that might arise, for example, when the centre of mass of a macroscopic body is prepared in a quantum mechanical state.

\section{Problems}
\paragraph*{Problem:} Taking the Hamiltonian (\ref{sarh:3dVham}) show that in the limit of a very massive body in uniform motion it becomes\\
\begin{equation}
	H=-\vec{V}\vec{\cdot}\int d^{3}\vec{x}\int_{0}^{\infty}d\omega\left[(\vec{\nabla}\vec{\otimes}\vec{X}_{\omega})\vec{\cdot}\vec{\Pi}_{\vec{X}_{\omega}}-\alpha(\omega,\vec{x}-\vec{R})\vec{X}_{\omega}\vec{\times}\vec{B}\right]+H_{0}+\text{const.}\label{sarh:probeqham}
\end{equation}
where \(\vec{V}=\vec{p}/M\).  Show that if the constant is neglected then this Hamiltonian can have an arbitrarily low energy (which could become negative), while the original Hamiltonian (\ref{sarh:3dVham}) is always positive.  What is the resolution of this apparent contradicition?\\

\paragraph*{Problem:} Find the equations of motion for the electromagnetic field and the reservoir from the Hamiltonian (\ref{sarh:probeqham}), and find expressions for the electromagnetic field in terms of the \(\vec{C}_{\omega}(k_{x},y,z)\) and \(\vec{C}_{\omega}^{\star}(k_{x},y,z)\) of the reservoir.  Assume that the medium is homogeneous and moves along the \(x\)--axis.\\

\paragraph*{Problem:} Extend the solutions of the above problem to the quantum case and show that a suitable Hamiltonian to describe the evolution of the operators is
\[
	\hat{H}=\int\frac{d k_{x}}{2\pi}\int_{0}^{\infty}d\omega\hbar(\omega+Vk_{x})\hat{\vec{C}}_{\omega}^{\dagger}(k_{x},y,z)\vec{\cdot}\hat{\vec{C}}_{\omega}(k_{x},y,z).
\]
What are the eigenvalues of this Hamiltonian and how do they differ from the \(V=0\) case?\\

\paragraph*{Problem:} Consider a detector coupled to the electromagnetic field with a Hamiltonian
	\[
		\hat{H}_{D}=\hbar\omega_{0}\left[\hat{a}^{\dagger}\hat{a}+\frac{1}{2}\right]
	\]
	and an interaction
	\[
		\hat{H}_{I}=i\beta\left[\hat{a}-\hat{a}^{\dagger}\right]\vec{e}\vec{\cdot}\hat{\vec{E}}(\vec{x}_{0},t),
	\]
	where \(\beta\) is a constant determining the strength of the interaction, and \(\vec{e}\) is a unit vector determining its orientation.  Suppose that the detector is initially in the first excited state and find an expression (valid to first order in perturbation theory) for the transition rate into the ground state in terms of the electric field operator.  Using the expression for the electric field operator given in the text evaluate this in terms of the Green function.\label{sarh:rateprob}\\

\paragraph*{Problem:}  Find an expression for the electromagnetic Green function in a region of space outside of a dielectric half space (\(x>0\)) characterised by reflection coefficients \(r_{\lambda}(\omega,\vec{k})\) for the two polarisations \(\lambda=1,2\).  Use the result of problem (\ref{sarh:rateprob}) to find the transition rate of the detector as a function of the distance from the surface.\\[10pt]
	\emph{Hint}: Start from the free space Green function, decompose it into a sum of freely propagating waves, and add in the reflected waves necessary to fulfil the boundary conditions.\\

\paragraph*{Problem:} Find an expression for the electromagnetic Green function in a region of space outside of a dielectric cylinder of radius \(R\) characterised by some \(\varepsilon(\omega)\).  Use the result of problem (\ref{sarh:rateprob}) to find the transition rate of the detector as a function of the distance from the cylinder.	 

\end{document}